\documentclass[pra,aps,superscriptaddress]{revtex4}
\usepackage{epsfig,amsmath,amssymb}
\begin{document}

\title{Relaxation and Metastability in the RandomWalkSAT search procedure}

\author{Guilhem Semerjian}
\email{guilhem@lpt.ens.fr}
\affiliation{CNRS-Laboratoire de Physique Th{\'e}orique
de l'ENS, 24 rue Lhomond, 75005 Paris, France.}
\author{R\'emi Monasson}
\email{monasson@lpt.ens.fr}
\affiliation{CNRS-Laboratoire de Physique Th{\'e}orique
de l'ENS, 24 rue Lhomond, 75005 Paris, France.}
\affiliation{CNRS-Laboratoire de Physique Th{\'e}orique, 
3 rue de l'Universit\'e, 67000 Strasbourg, France.}

\date{\today}

\begin{abstract}
An analysis of the average properties of a local search resolution
procedure for the satisfaction of random Boolean constraints is
presented. Depending on the ratio $\alpha$ of constraints per
variable, resolution takes a time $T_{res}$ growing linearly ($T_{res}
\sim \tau_{res}(\alpha) \, N, \alpha < \alpha _d$) or exponentially
($T_{res} \sim \exp( N\, \zeta(\alpha)), \alpha > \alpha _d$)
with the size $N$ of the instance. The relaxation time $\tau_{res}(\alpha)$
in the linear phase is calculated through a systematic expansion scheme
based on a quantum formulation of the evolution operator. For 
$\alpha > \alpha_d$, the system is trapped in some metastable state, and 
resolution occurs from escape from this state through crossing of a 
large barrier. An annealed calculation of the height $\zeta (\alpha)$ 
of this barrier is proposed. The polynomial/exponentiel cross-over
$\alpha _d$ is not related to the onset of clustering among solutions.
\end{abstract}

\preprint{LPT-ENS 02/66}

\maketitle

\section{Introduction}

The study of combinatorial problems\cite{papa} with statistical
physics techniques started almost twenty years ago\cite{revue}. Most
of the efforts have been devoted to the calculation of the optimal
solution of various problems (traveling salesman, matching, graph or number
partitioning, satisfiability of Boolean constraints, vertex cover of
graphs...) as a function of the definition parameters of their
inputs distributions. Central to these studies is
the characterization of the properties of the extrema of correlated random
variables, a question of considerable importance in probability
theory\cite{tala}.  From a computer science point of view, however,
the main point is the characterization of resolution times. Concepts
and tools issued from the analysis of algorithms have allowed so far
to understand the behaviour and the efficiency of many algorithms of
practical use~\cite{Knu,flaj}, sorting for instance, but
progress has been much slower in the analysis of search procedures
for combinatorial problems. These are sophisticated algorithms hardly
amenable to rigorous analysis with available techniques\cite{Fra}.
Statistical physics ideas and approximation techniques may then be of
great relevance to help develop quantitative understanding and
intuition about the operation of these algorithms. To some extent, the
study of algorithms may be seen as part of out-of-equilibrium
statistical physics.

There exists a wide variety of algorithms for combinatorial
problems\cite{papa}.  Roughly speaking, two main classes may be
identified. The first one includes complete algorithms, guaranteed to
provide the optimal solution. They essentially
consist in an exhaustive albeit clever (making use of branch--and--bound
procedures) search through the configuration space, and may require
very large computational times, {\em i.e.} scaling exponentially with
the size of the inputs to be treated.  Recently, notions borrowed from
statistical physics as real-space renormalization and
out-of-equilibrium growth processes allowed to reach some
understanding of the operation of complete algorithms for the
satisfiability\cite{coc1,coc2} and the vertex cover\cite{mart,andr}
problems over random classes of inputs.  Incomplete algorithms
constitute another large class of resolution procedures: they may be
able to find the optimal solution very quickly, but may also run
forever without ever finding it. An example is provided by local
procedures e.g. Monte Carlo dynamics which attempt at finding a
solution from an arbitrary initial configuration through a sequence 
of stochastic local moves in the configuration space. When
specialized to decision problems (for which the desidered output is
the answer YES or NO to a question related to the inputs, as: is there
a way to color a given graph with 7 colors only?), local search
algorithms can sometimes be made one-sided error probabilistic
algorithms\cite{rand}.  When they stop, the answer is YES with
certainty.  If they run for a time $t$ without halting, the
probability (over the non deterministic choices of local moves in the
configuration space) that the correct answer is NO is bounded from
below by a function $f(t,N)$ going to 1 when $t\to\infty$. Obviously
this function depends on the size $N$ of the input: the larger the
input, the larger the time $t_c(N)$ it takes to reach, say, a 99\%
confidence that the answer is NO {\em i.e.} $f(t_c(N),N)=0.99$.  
Determining the scaling (polynomial
or exponential) of $t_c(N)$ with $N$ is of capital importance to
assess the efficiency of the algorithm.

In this paper, we study this question for a local search procedure,
the RandomWalkSAT algorithm\cite{papa2}, and a decision problem defined over
an easy-to-parametrize class of inputs, the random satisfiability
problem\cite{mitc}. Both procedure and problem are defined in Section~\ref{sec_def}. We also recall the main results proven by mathematicians on these
issues, and present an overview of the phenomenology of RandomWalkSAT. A
major theoretical interest for studying the RandomWalkSAT algorithm is that
it is, apparently, of purely dynamical nature.  Detailed balanced is
indeed verified in a trivial way: the equilibrium measure is non zero over
solutions only, and the transition rates from a solution to any 
other configuration are null.
The equilibrium measure is therefore of no use to understand long
time dynamics, a situation
reminiscent of some models studied in out-of-equilibrium physics
e.g. the contact process for finite size systems\cite{ligg}.  We are
thus left with a study of the dynamical evolution of a spin system
with disorder in the interactions (the instances of the combinatorial
problem to be solved are random), a still largely open problem in
statistical physics\cite{leti}. We show in Section~\ref{sec_exact}
 how the master
equation for this evolution can be written as a Hamiltonian (in
imaginary time) for 1/2 quantum spin systems, and use this
represention to get exact results and develop systematic expansions
for the quantities of interest, valid in some region of the parameter
space.  In Section~\ref{annealed_sec}, 
we present an approximate analysis of the
RandomWalkSAT procedure in the whole parameter space. We show that,
depending on the value of the ratio $\alpha$ of the number of
constraints per (Boolean) degree of freedom, resolution is either
achieved in linear time or requires the escape from some metastable
region in the configuration space, a slow process taking place over
exponentially large times. Interestingly, the dynamics
generated by RandomWalkSAT is very similar to the physical dynamics of
(spin) glassy systems\cite{leti}.  Some perspectives are presented in
Section~\ref{sec_conclu}. 
Note that a complementary study of the RandomWalkSAT procedure 
was very recently carried out by Barthel, Hartmann and Weigt\cite{and}.

\section{Definitions, known results and phenomenology}
\label{sec_def}
\subsection{The random $K$-Satisfiability problem}

The 3-Satisfiability (3-SAT) decision problem is defined as follows.
Consider a set of $N$ boolean variables $x_i$, $i=1,\ldots ,N$. A
literal is either a variable $x_i$ or its negation $\bar{x}_i$. A clause
is the logical OR between 3 distinct literals. It is thus true as soon
as one of the literals is true. A formula is the logical AND between
$M$ clauses, it is true if and only if all the clauses are true. 
A formula is said to be {\em satisfiable} if there is an assignment of
the variables such that the formula is true, {\em unsatisfiable}
otherwise. 

3-SAT is a NP-complete problem\cite{papa}; it is believed
that there is no algorithm capable of solving every instance of 3-SAT in
a time bounded from above by a polynomial of the size of the instance. 
How well do existing and {\em a priori} exponential algorithm perform
in practice? To answer these questions, computer scientists have
devised a simple way of generating random instances of the 3-SAT problem,
with a rich pattern of hardness of resolution\cite{mitc}. 
Formulas are drawn in the following way. Repeat $M$ times
independently the same process: pick up a $3$-uplet of distinct
indices in $[1,N]$, uniformly on all possible $3$-uplets. For each of
the $3$ corresponding variables, choose the variable itself or its 
negation with equal probability ($1/2$), and construct a clause with 
the chosen literals. Repetition of this process $M$ times gives a set
of $M$ independently chosen clauses, whose conjunction is the generated 
instance. The random generation
of formula makes the set of possible formulas a probability space,
with a well defined measure. 

Numerical experiments indicate that a phase transition takes 
place when $N,M \to \infty$ at fixed ratio $\alpha = M/N$ of clauses 
per variables (in the thermodynamic limit).
If $\alpha$ is smaller than some critical value 
$\alpha _c \simeq 4.3$, a randomly drawn instance admits at least one
solution with high probability. Beyond this threshold, instances are
almost never  satisfiable. The existence of this transition has not been 
proven rigorously yet\cite{frie}, 
but bounds on the threshold exist: the probability
of satisfaction tends to 1 (respectively to 0) if $\alpha < 3.42$\cite{kiro}
(resp. $\alpha > 4.506$\cite{dubo})
All the above definitions can be extended to the $K$-SAT problem, where
each clause is the disjunction of $K$, rather than  $3$, literals.
2-SAT is an easy (polynomial) problem, while $K$-SAT is NP-Complete
for any $K\ge 3$. Location of the threshold is rigorously known for 
2-SAT ($\alpha _c=1$)\cite{Goerdt} but
not for $K\ge 3$. We shall denote in the following averages on the random $K$-SAT ensemble by $[\cdot]$. 

Statistical mechanics studies have pointed out the existence of
another phase transition taking place in the satisfiable phase
($\alpha < \alpha_c$) with a location, first estimated to $\alpha_s 
\simeq 3.95$\cite{giul} and then to  $\alpha_s \simeq 
3.92$\cite{meza}. This phase transition is related to the microscopic
structure of the set of solutions. Define $d$ the Hamming distance
between two solutions as the number of variables taking opposite
values in these solutions.  When $\alpha < \alpha _s$, 
there exist an exponentially large (in $N$) number of solutions, each
pair of which are separated by a path in the solution space, that is,
a sequence of solutions with a $O(1)$ Hamming distance between
successive solutions along the path. The solution space is made of a
single cluster of solutions. For $\alpha _s < \alpha < \alpha _c$, the
solution space breaks into an exponentially large (in $N$) number of
clusters, separated by large voids prived of solutions.  Two solutions
in one cluster are linked through a path while there is no path in the
solution space linking two solutions in two different clusters. This
clustering phenomenon, whose discovery was inspired from previous
works in the context of information storage in neural
networks\cite{net}, was subsequently found in various combinatorial
problems\cite{xorsat1,gc}, and rigorously demonstrated for the
so-called random XORSAT problem\cite{xorsat2}.  It is a zero
temperature signature of the ergodicity breaking taking place in spin
glasses\cite{leti,sg}.  Its precise relationship with dynamical
properties, and in particular with the computational cost for finding
a solution is not fully elucidated yet\cite{giul,suedois,meza}. 

\subsection{The RandomWalkSAT algorithm} 

The operation of RandomWalkSAT (also called RandomWalk) on an instance of the $K$-SAT problem is as follows\cite{papa2}.
\begin{enumerate}
\item Choose randomly a configuration of the Boolean variables. 
\item If all clauses are satisfied, output ``Satisfiable''. 
\item If not, choose randomly one
of the unsatisfied clauses, and one of the $K$ variables of this
clause. Flip the variable.
Notice that the selected clause is now satisfied, but the flip
operation may have violated other clauses which were previously satisfied. 
\item Go to step 2, until a limit on the number of flips fixed beforehand 
has been reached. Then Output ``Don't know''.
\end{enumerate}
 
What is the output of the algorithm? Either ``Satisfiable'' and a
solution is exhibited, or ``Don't know'' and no
certainty on the status of the formula is reached.  Papadimitriou
showed that RandomWalkSAT solves with high probability any satisfiable
2-SAT instance in a number of steps (flips) of the order of
$N^2$\cite{papa2}. Recently Sch\"oning was able to prove the following
very interesting result for 3-SAT\cite{scho}. Call `trial' a run of
RandomWalkSAT consisting of the random choice of an intial
configuration followed by $3\times N$ steps of the procedure. If none
of $T$ successive trials of RandomWalkSAT on a given instance has been
successfull (has provided a solution), then the probability that this
instance is satisfiable is lower than $\exp( - T \times (3/4)^N)$. In
other words, after $T\gg (4/3)^N$ trials of RandomWalkSAT, most of the
configuration space has been `probed': if there were a solution, it
would have been found.  Though RandomWalkSAT is not a complete
algorithm, the uncertainty on its output can be made as small as
possible and it can be used to prove unsatisfiability (in a
probabilistic sense).

Sch\"oning's bound is true for any instance. Restriction to special
input distributions allows to improve his result. Alekhnovich and
Ben-Sasson showed that instances drawn from the random
3-Satisfiability ensemble described above are solved in polynomial
time with high probability when $\alpha$ is smaller than
$1.63$\cite{ben}. It is remarkable that, despite the quenched
character of the disorder in this problem (the same clauses are seen
various times in the course of the search), rigorous results on the
dynamics of this spin model can be achieved.

\subsection{Phenomenology of the operation of RandomWalkSAT}
\label{sub_phen}
\begin{center}
\begin{figure}
A\epsfig{file=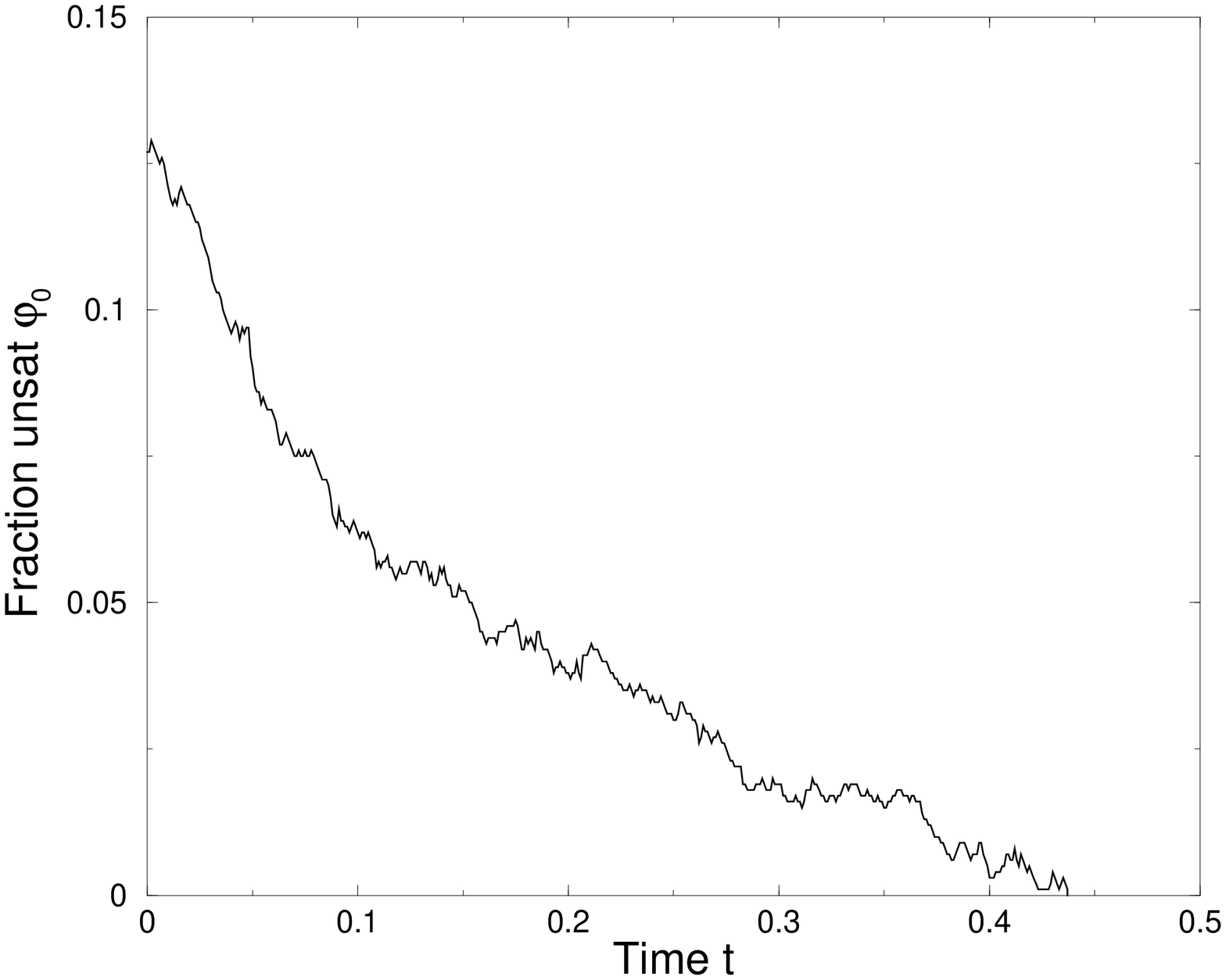,angle=-90,width=7cm}
\hskip 1cm
B\epsfig{file=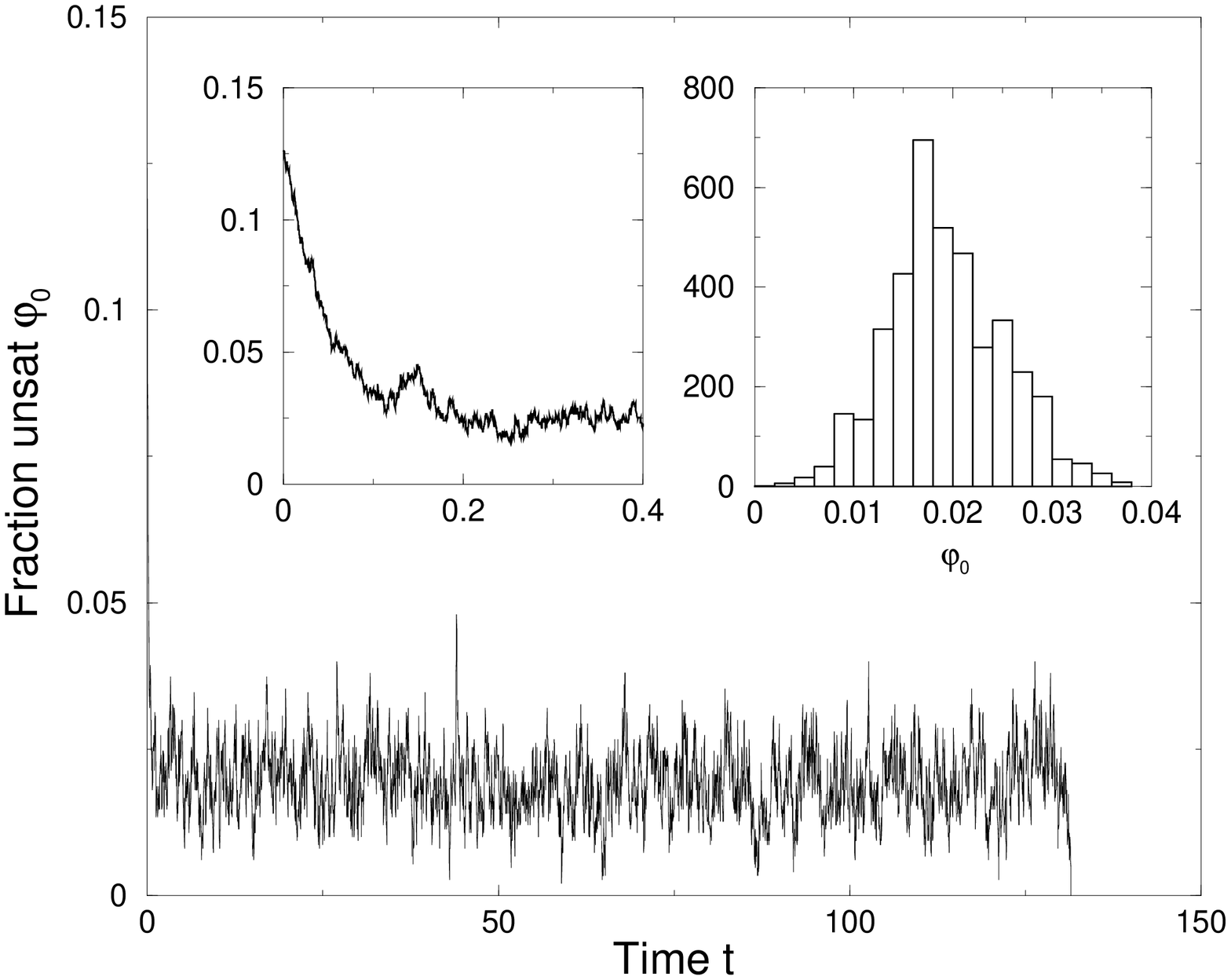,angle=-90,width=7cm}
\vskip .5cm
\caption{Fraction $\varphi _0$ of unsatisfied clauses as a function of time
$t$ (number of flips over $M$) for two randomly drawn instances of 3-SAT with
ratios $\alpha =2$ ({\bf A}) and $\alpha =3$ ({\bf B}) with $N=500$ variables. 
Note the difference of time scales between the two figures. Insets of
figure B: left: blow up of the initial relaxation of $\varphi_0$, taking place
on the $O(1)$ time scale as in ({\bf A}); right: histogram $p_{500}
(\varphi _0 )$ of the 
fluctuations of $\varphi _0$ on the plateau $1\le t\le 130$. }
\label{phen}
\end{figure}
\end{center}

\begin{center}
\begin{figure}
A\epsfig{file=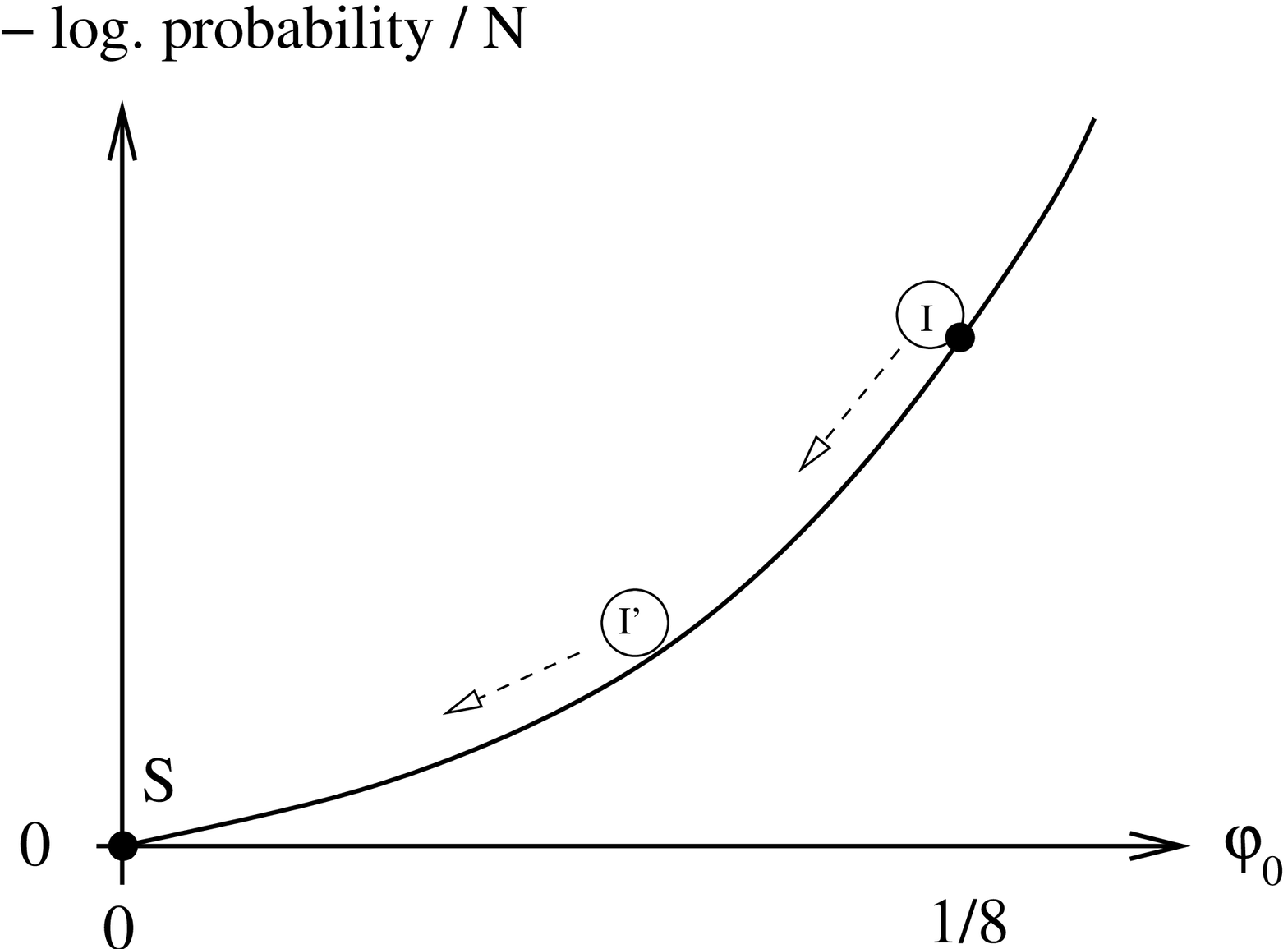,angle=0,width=7cm}
\hskip 1cm
B\epsfig{file=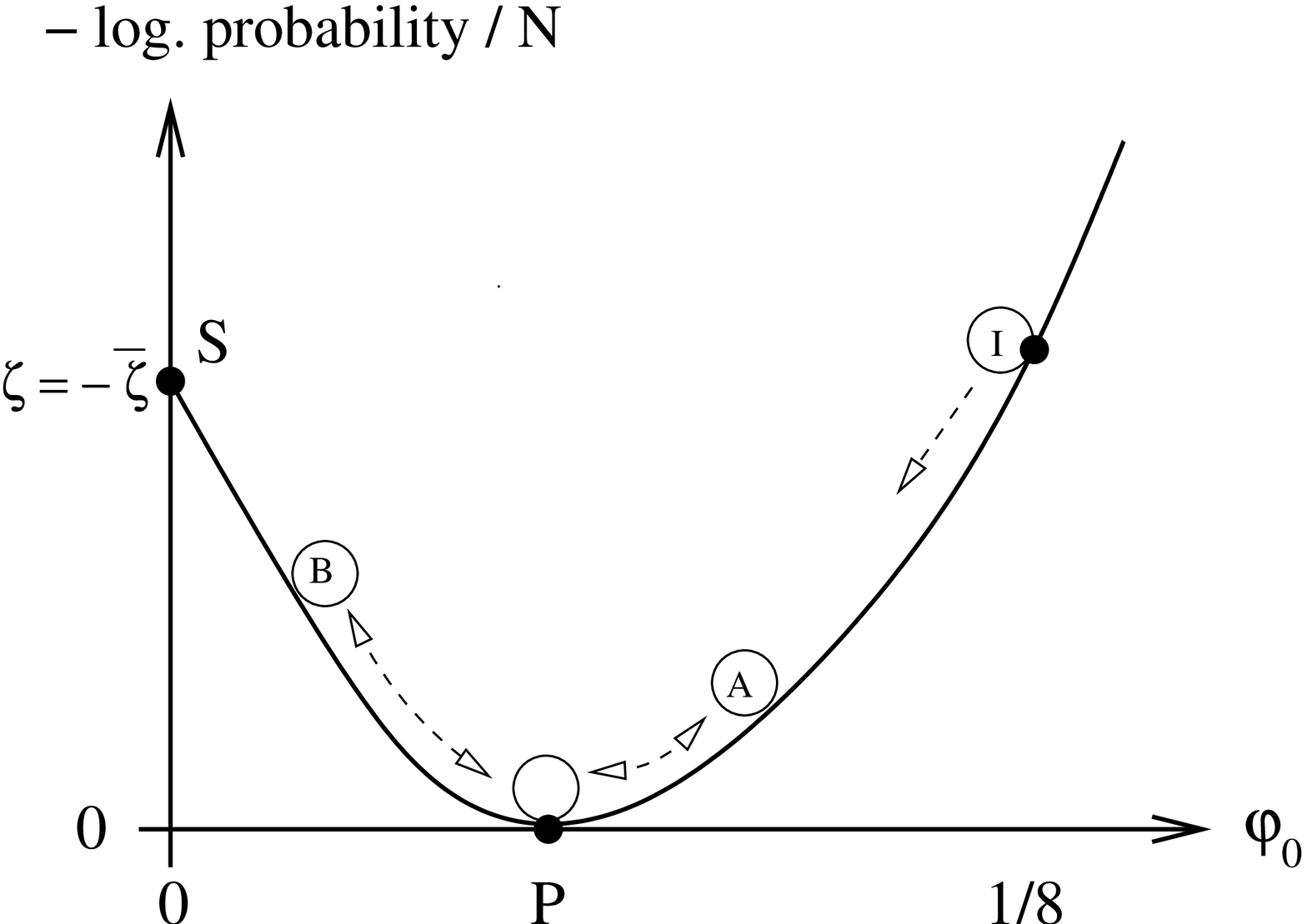,angle=0,width=7cm}
\vskip .5cm
\caption{Schematic picture of the operation of RandomWalkSAT
for ratios $\alpha$ smaller ({\bf A}) or larger
({\bf B}) than the dynamical threshold $\alpha _d$. Vertical axis is
minus the logarithm of the probability (divided by $N$) that the system
has a fraction $\varphi_0$ of unsat clauses after a large number of
RandomWalkSAT flips. Its representative curve can be seen as an energy
potential in which the configuration (represented by an empty ball) 
rolls down towards the more probable value of the order parameter $\varphi _0$,
or up, through stochastic fluctuations. The starting configuration violates
$\varphi _0 =1/8$ of the clauses (point I). 
At small ratios ({\bf A}), the configuration
rolls down to reach point S through a sequence of intermediary
points (I'). Resolution is essentially a fast relaxation towards S
($O(1)$ time scale).
At large ratios ({\bf B}), the ball first relaxes to the bottom
of the well (point P with abscissa corresponding to the plateau height).
Then slow negative (ball A) or positive (ball B) fluctuations of the
fraction $\varphi _0$ take place on exponentially (in $N$) long time 
scales. The time $t_{res} \sim \exp ( N \, \zeta)$ it takes to
the system to reach a solution (point S) is, on the average, equal to
the inverse probability $\exp( N \, \bar \zeta)$ that a fluctuation
drives the system to S.}
\label{schema}
\end{figure}
\end{center}

In this section, we briefly sketch the behaviour of RandomWalkSAT, as
seen from numerical experiments~\cite{Parkes} and the analysis 
presented later in
this article.  We find that there is a dynamical threshold $\alpha _d$
separating two regimes:
\begin{itemize}
\item for $\alpha < \alpha _d \simeq 2.7$ for 3-SAT, the algorithm
finds a solution very quickly, namely with a number of flips
growing linearly with the number of variables $N$.  Figure~\ref{phen}A
shows the plot of the fraction $\varphi _0$ of unsatisfied clauses as
a function of time $t$ (number of flips divided by $M$) for one
instance with ratio $\alpha=2$ and $N=500$ variables. The curves show
a fast decrease from the initial value ($\varphi _0(t=0)=1/8$ in the
large $N$ limit independently of $\alpha$) down to zero on a time
scale $t_{res}=O(1)$. Fluctuations are smaller and smaller as $N$ grows.
$t_{res}$ is an increasing function of $\alpha$.  This {\em relaxation} regime
corresponds to the one studied by 
Alekhnovich and Ben-Sasson, and $\alpha _d > 1.63$
as expected\cite{ben}. Figure~\ref{schema}A symbolizes the behaviour
of the system in the relaxation regime.

\item for instances in the $\alpha_d < \alpha  < 
\alpha _c$ range, the initial relaxation phase taking place
on $t=O(1)$ time scale is not sufficient to reach a solution
(Figure~\ref{phen}B). The fraction $\varphi_0$ of unsat clauses then 
fluctuates around some plateau value for a very long time. 
On the plateau, the system is trapped in a {\em metastable} state.
The life time of this metastable state (trapping time) is so huge that it is
possible to define a (quasi) equilibrium probability distribution 
$p_N(\varphi _0)$ for the fraction $\varphi_0$ of unsat clauses.
(Inset of Figure~\ref{phen}B). The distribution of fractions is well 
peaked around some average value (height of the plateau),
with left and right tails decreasing exponentially fast with $N$,
$p_N(\varphi _0) \sim \exp ( N \bar \zeta (\varphi_0))$
with $\bar \zeta \le 0$ (Figure~\ref{schema}B).
Eventually a large negative fluctuation will bring the system to
a solution ($\varphi _0=0$). Assuming that these fluctuations are independant random events occuring with probability $p_N(0)$ on an interval of time of order $1$, the resolution time is a stochastic variable with exponential distribution. Its average is, to leading exponential order, 
the inverse of the probability of resolution on the $O(1)$ time scale: $[t_{res}] \sim
\exp (N \zeta)$ with $\zeta = - \bar \zeta (0)$.
Escape from the metastable state therefore occurs through
barrier crossing and 
takes place on exponentially large--in--$N$ time scales, as confirmed
by numerical simulations for different sizes. 
Sch\"oning's result\cite{scho} can be interpreted as a lower bound to the
probability $\bar \zeta (0)> \ln (3/4)$, true for any instance. 
\end{itemize}

\begin{center}
\begin{figure}
\epsfig{file=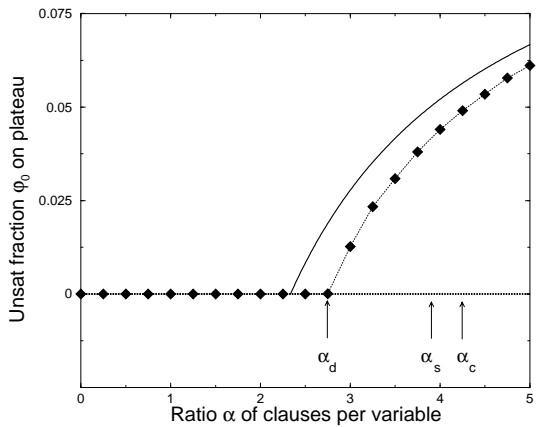,angle=-90,width=7cm}
\vskip .5cm
\caption{Fraction $\varphi _0$ of unsatisfied clauses on 
the metastable plateau as
a function of the ratio $\alpha$ of clauses per variable. Diamonds  
are the output of numerical experiment, and have
been obtained through average of data from simulations at a given
size $N$ (nb. of variables) over 1,000 samples of 3-SAT, and extrapolation
to infinite sizes (dotted line serves as a guide to the eye). 
The ratio at which $\varphi _0$ begins being
positive, $\alpha _d \simeq 2.7$, is smaller than the thresholds
$\alpha _s\simeq 3.9$ and $\alpha _c\simeq 4.3$ above which solutions
gather into distinct clusters and instances have almost surely
no solution respectively. 
The full line represents the prediction of the Markovian
approximations of Section~\ref{sub_annealC} and \ref{sub_annealE}.}
\label{plateau}
\end{figure}
\end{center}

The plateau energy, that is, the fraction of unsatisfied clauses
reached by RandomWalkSAT on the linear time scale is plotted on
Figure~\ref{plateau}.  Notice that the ``dynamic'' critical value
$\alpha_d$ above which the plateau energy is positive (RandomWalkSAT
stops finding a solution in linear time) is strictly smaller than the
``static'' ratio $\alpha_c$, where formulas go from satisfiable with
high probability to unsatisfiable with high probability.  In the
intermediate range $\alpha_d < \alpha <\alpha_c$, instances are almost
surely satisfiable but RandomWalkSAT needs an exponentially large time
to prove so.  Interestingly, $\alpha _d$ and $\alpha_c$ coincides for
2-SAT in agreement with Papadimitriou's result\cite{papa2}. Furthermore,
the dynamical transition is apparently not related to the onset
of clustering taking place at $\alpha _s$.

\section{Exact results: special cases and expansions}
\label{sec_exact}
\subsection{Evolution equations and quantum formalism}

Boolean variables will be hereafter represented by Ising spins,
 $S_i = 1$ (resp. $-1$) when the Boolean variable $x_i$ is true
(resp. false). A microscopic configuration $\bf S$ is specified by 
the states of all variables: ${\bf S}=(S_1, S_2 , \ldots , S_N)$. 
We then define a $2^N$ dimensional linear space with canonical basis 
$\{|{\bf S}\rangle \}$, orthonormal for the scalar product $\langle{\bf S'}|{\bf S}\rangle=\prod_i \delta_{S_i',S_i}$.
Let us denote $\mbox{Prob}[{\bf S},T]$ the probability
that the system configuration is $\bf S$ at time $T$ {\em i.e.}
after $T$ steps of the algorithm, and define~\cite{KadaSwift}
\begin{equation} \label{sv}
|{\bf S}(T)\rangle = \sum_{{\bf S}} \mbox{Prob}[{\bf S},T]\; |{\bf
 S}\rangle
\end{equation}
as the (time-dependent) vectorial state of the system. 
Knowledge of this vector gives access to the probability 
$\mbox{Prob}[{\bf S},T]=\langle{\bf S}|{\bf S}(T)\rangle$
of being in a certain configuration $\bf S$.

RandomWalkSAT defines a Markov process on the set of configurations;
during one step of the algorithm the state vector of the system changes 
according to 
\begin{equation} \label{evolu1}
|{\bf S}(T+1)\rangle = \hat{W}_d \; |{\bf S}(T)\rangle
\end{equation}
where the evolution operator in discrete time $\hat{W}_d$ reads
\begin{equation} \label{discop}
\hat{W}_d = \sum _{\ell =1} ^M \hat F_\ell \cdot \hat U_\ell \cdot \hat E ^{-1} 
\ , \quad 
\hat F _\ell = \frac{1}{K} \sum_{i=1}^N C_{li}^2 \sigma _i ^x
 \ , \quad 
\hat U _\ell =  g_K \left(\sum_{j=1}^N C_{lj} \sigma _j
^z\right) \ , \quad
\hat E = \sum_{\ell = 1}^M \hat U _\ell \ .
\end{equation}
In the above expression, we have made use of several new notations that
we now explain. 
The $M \times N$ matrix $C_{\ell i}$ encodes the instance :
$C_{\ell i}$ equals $1$ if the $\ell$-th clause contains the literal $x_i$,
$-1$ if it contains the literal $\bar{x}_i$, and $0$
otherwise. Since every clause contains $K$ literals, 
$\sum_i C_{\ell i}^2=K \; \forall \ell$. 
The Pauli operators are defined through $\sigma _i ^z |{\bf
S}\rangle = S_i |{\bf S}\rangle$, and $\sigma _i
^x |{\bf S}\rangle = |{\bf S}^i\rangle$, where
${\bf S}^i$ is the configuration obtained from ${\bf S}$ by flipping
the $i$-th spin. It is a simple check that the argument of function $g_K$ 
in $\hat U_\ell$ is a diagonal operator in the canonical basis 
$\{|\bf S\rangle\}$, with eigenvalues 
$x\equiv\sum_{i=1}^N C_{\ell i} S_i$  in
$\{-K,-K+2,\dots,K-2,K\}$, equal to $-K$ if and only if 
 clause $\ell$ is unsatisfied.
The function 
\begin{equation}
g _K (x) = \delta_{x,-K} = \frac 1{2^K\, K!}\prod _{p=0} ^{K-1}
\left( { K-2p - x } \right)
\end{equation}
is a polynomial of degree $K$ in $x$ equal to 1 if $x=-K$ and to zero
for all the other possible eigenvalues.  Operator $\hat{E}$ is diagonal
in the canonical basis too, with eigenvalues equal to the numbers of
unsatisfied clauses (also called energy) in each configuration.
Therefore $\hat U _\ell \cdot
\hat E^{-1}$ in (\ref{discop}) acts as a filter retaining clause $\ell$
only if it unsatisfied, and with a  weight equal to the inverse number of
unsatisfied clauses: only unsat clauses can be
chosen at each time step, each of them with the same probability.
$\hat F _\ell$ flips the spins of clause $\ell$, each with
probability $1/K$. Notice that $\hat{W}_d$ conserve
probabilities:  $\langle O | \hat{W}_d=\langle O |$ where 
$\langle O| = \sum_{\bf S} \langle {\bf S}|$ is the superposition
of all possible states.

In the thermodynamic limit, the evolution can be rewritten in
continuous time, defining $t=T/M$,
\begin{equation}
\frac{d}{dt} |{\bf S }(t)\rangle = \hat{W} |{\bf S}(t)\rangle 
\qquad , \qquad \qquad \hat{W} = M(\hat{W}_d - \hat{1})
\end{equation}
Formally, the solution of this equation is $|{\bf S}(t)\rangle =
e^{t\, \hat{W} } |{\bf S}(0)\rangle$, where $|{\bf S}(0)\rangle=(1/2^N)
\sum_{\bf S} |{\bf S} \rangle$ since the initial configuration is
random.  An important quantity to compute is the average fraction of
unsatisfied clauses at time $t$, {\em i.e.} after $T= M t$ steps of the
algorithm, averaged both on the history of the algorithm and on the
distribution of formulas:
\begin{equation}
\varphi_0(t) = \frac{1}{M} \left[\langle O | \hat{E} e^{t\,\hat{W}}
|{\bf S}(0)\rangle \right]
\end{equation}
For the sake of analytical simplicity, we shall study a slightly
different evolution operator in the next two subsections, and use the
parameter $u$ to denote the time parameter of this modified evolution,
\begin{equation}
\hat{W}'= \hat{W} \cdot \frac{1}{M} \hat{E} = \sum_{\ell=1}^M
\left(\hat{F}_\ell - \hat{1} \right)\cdot \hat{U}_{\ell} \quad ,
\qquad \frac{d}{du} |{\bf S} (u)\rangle = \hat{W}' |{\bf S} (u)\rangle
\label{def_Wp}
\end{equation}
Let us explain the meaning of this modification. The operator
$\hat{W}'$ would have been obtained starting from the following
variant of the RandomWalkSAT stochastic process. At each step, choose a
clause among the $M$ ones. If it is satisfied, do nothing. If
it is unsatisfied, flip one of its variables. 

In the thermodynamic 
limit, the fraction of unsatisfied clauses $\varphi _0$ is expected 
to become a self averaging quantity, {\em i.e} to be peaked with high 
probability around its ($t$-dependent) mean value. 
Turning $\hat{W}$ into $\hat{W}'$ thus amounts to a local redefinition 
of time, independent of the instance of the problem. In 
definition~(\ref{def_Wp}), the operator $\hat{E}/M$ can be 
replaced with its mean value
$\varphi_0$, leading to $\frac{d}{du} \equiv \varphi_0 \frac{d}{dt}$. 
The knowledge of the $\hat W'$--evolution of any observable in terms 
of $u$ can then be rewritten in terms of $t$ through
\begin{equation}
t(u) = \int_0^u du' \varphi_0(u') \qquad .
\label{eq_rescaling}
\end{equation} 
Examples of this time reparametrization will be shown below.

\subsection{The $K=1$ case}

Let us first study the simple case $K=1$. A clause is then a single
literal, {\em i.e.} either a variable or its negation. If both a variable
and its negation appear in the formula, it is obviously
unsatisfiable. This is the case, with high probability in the
thermodynamic limit, as soon as $\alpha>0$. Static properties of 
1-SAT model are known
exactly~\cite{mz}, and its dynamics under the
RandomWalkSAT evolution can  be solved too.

\begin{center}
\begin{figure}
\epsfig{file=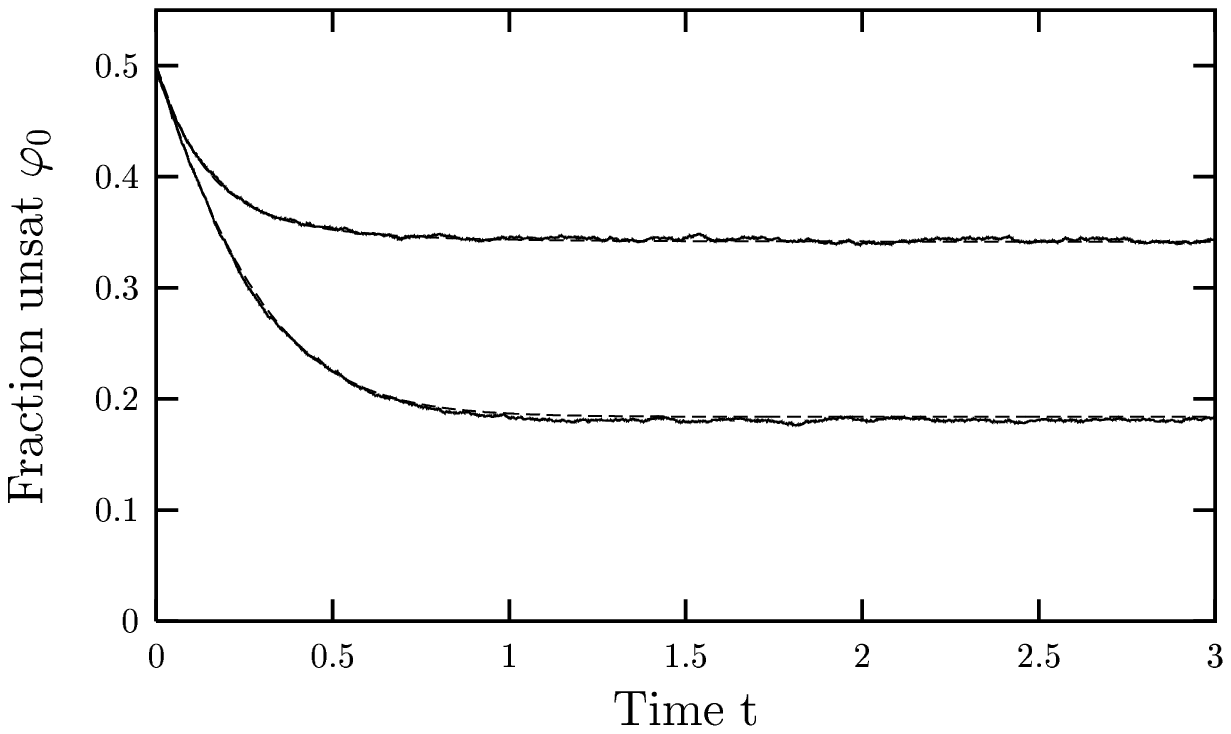,width=220pt} \hspace{1cm}
\epsfig{file=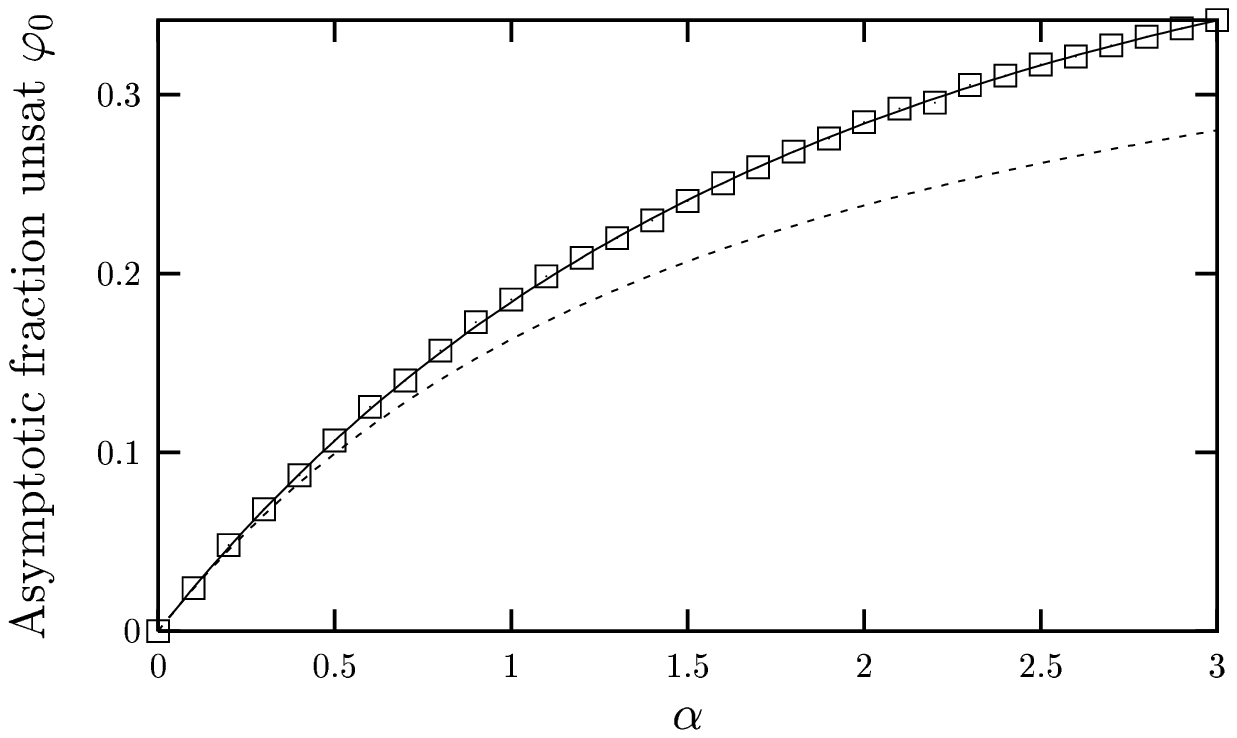,width=220pt}
\caption{Comparison between numerics and analytical results for $K=1$. 
Left panel: fraction of unsat clauses $\varphi_0(t)$ as a function of time 
$t$ for $\alpha=3$ (top) and $\alpha=1$ (bottom). 
Dashed line: analytical curve, solid line (almost superimposed): numerical results (single run with $N=10^4$). 
Right panel: asymptotic fraction of unsat clauses as a function of 
$\alpha$, solid line: theoretical prediction (\ref{asty}), 
symbols: numerical results (averaged over 10 runs, over $t\in [8,10]$, 
for $N=10^4$), dashed line: groundstate energy $\varphi_{GS}$.}
\label{fig_eneK1}
\end{figure}
\end{center}

An instance is described by a set of integers $\{m_i,n_i\}$, 
where $m_i$ is the number of clauses in which the variable $i$ appears, 
and $(m_i - n_i)/2$ is the number of times it has been chosen without negation.
The evolution operator in terms of $u$--time is the sum of site-operators 
\begin{equation}
\hat{W}_i' = \frac{1}{2} \left( m_i \sigma_i^x - m_i + n_i
\sigma_i^z - n_i \sigma_i^x\sigma_i^z \right)
\end{equation}
As the operators $\hat{W}_i'$ on different sites commute, the
evolution operator can be diagonalized in each of the $i$ subsets
independently, and the vector state is a tensor product,
\begin{equation}
|{\bf S}(u)\rangle = \bigotimes_{i=1}^N
\left\{ \begin{array}{c c}
\frac{1}{2} (|+\rangle_i + |-\rangle_i) & \mbox{if} \; m_i=0 \\
\frac{1}{2} \sum_{S=\pm 1} (1 + S \frac{n_i}{m_i} (1-e^{-m_i
u})) |S\rangle_i & \mbox{if} \; m_i \neq 0
\end{array} \right.
\end{equation}
where $|\pm \rangle_i$ are the eigenvectors of $\sigma_i^z$ with 
eigenvalues $\pm 1$. The fraction of unsatisfied clauses for
a given instance and averaged 
over the choices of the algorithm reads,
\begin{equation}
\frac 1M \, \langle O|\hat{E}|{\bf S}(u) \rangle = 
\frac{1}{2} - \frac{1}{2M} \sum_{i /
m_i \neq 0} \frac{n_i^2}{m_i} \left( 1 - e^{- m_i u} \right)
\end{equation}
In the thermodynamic limit, the $m_i$'s become independent random 
variables with identical Poisson distributions of parameter $\alpha$; 
$(m_i-n_i)/2$ obeys a Binomial law of parameter $1/2$ among $m_i$.
Performing the quenched average over the formula,
the fraction of unsatisfied clauses reads
\begin{equation}
\varphi_0(u)= \frac{1}{2} -
\frac{1-\exp(\alpha(e^{-u} -1))}{2\alpha} 
\end{equation}
These exact results are compared with numerical simulations in
Figure~\ref{fig_eneK1}. On the right panel has been drawn the
asymptotic fraction of unsatisfied clauses obtained at large times,
compared with the analytical prediction made above. The left panel
shows the time evolution of the fraction of unsatisfied clauses as a
function of time $t$, for two different values of $\alpha$. The
analytical curve has been obtained through the rescaling explained at
the end of the previous subsection: from the exact value of
$\varphi_0(u)$ given above, the original time $t(u)$ has been obtained by
numerical integration (eqn.~(\ref{eq_rescaling})), and the plot
$\varphi_0(t)$ is a parametric plot $\{t=t(u),\varphi_0=\varphi_0(u)\}$,
parametrized by $u$.

Note that the asymptotic value of the energy reached is larger than
the groundstate one~\cite{mz},
\begin{equation} \label{asty}
\lim_{t \to \infty} \varphi_0(t) = \frac{1}{2} + \frac{e^{-\alpha}-1}{2
\alpha} > \frac 12 \big[ 1-e^{-\alpha} I_0(\alpha)-e^{-\alpha} I_1(\alpha)
\big]
= \varphi_{GS}
\end{equation}
where $I_n$ is the $n$-th modified Bessel function. This is 
easily understood: if a formula contains $x_1$ once and 
$\bar{x}_1$ twice, the optimal value is $x_1$ false. 
But the algorithm does not stop in this configuration, as the clause $x_1$ 
is then violated. RandomWalkSAT will keep on flipping the variable 
and make the average energy higher than the optimal one.

\subsection{The satisfiable phase of the $K=2$ case}
\label{sub_K2}
We now turn to the 2-SAT case, where every clause
involves two variables. It has been rigorously proven that there is a
sharp threshold phenomenon taking place at $\alpha_c=1$ in this
problem~\cite{Goerdt}: for lower values of $\alpha$ the formulas are
almost always satisfiable, beyond this threshold they are almost
always unsatisfiable. The dynamical critical threshold of
RandomWalkSAT has the same value, $\alpha _d= 1$\cite{papa2}. 
Thus there is no metastability in 2-SAT: if a solution of the formula 
exists it is almost surely found in polynomial
time. Two questions of interest are: how fast will the
number of unsatisfiable clauses decrease during the evolution of the process 
and how long will it take the algorithm to find a solution for $\alpha < 1$? 
What will be the
typical energy after a long run of the algorithm when $\alpha >1$? 
In this and the next subsections we shall address the first of
these two questions, and leave the second one to the next section.

\begin{center}
\begin{figure}
\epsfig{file=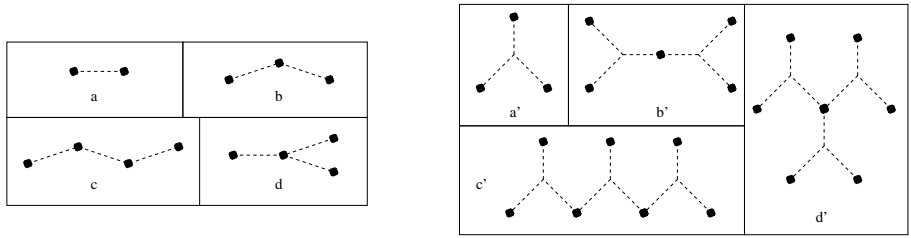,width=12cm}
\vskip .5cm
\caption{Tree-like clusters contributing to the expansions. Left panel: 
$K=2$. Right panel: $K=3$.}
\label{fig_clus}
\end{figure}
\end{center}

As in the $K=1$ case, one can use the quantum formalism and write down
the evolution operator $W'$. There appear both single site terms and 
couplings between pairs of sites,
\begin{eqnarray}
\hat{W}'&=&\sum_i\hat{W}'_i+\sum_{i \neq j} \hat{W}'_{ij} \\ \hat{W}'_i
&=& \frac{1}{8} \left(m_i \sigma_i^x - m_i +2 n_i \sigma_i^z -n_i
\sigma_i^x \sigma_i^z \right) \\ \hat{W}'_{ij} &=& \frac{1}{8}
\left[a_{ij} \left( \frac{\sigma_i^x + \sigma_j^x}{2} - 1 \right)
\sigma_i^z \sigma_j^z - b_{ij} \sigma_i^x \sigma_j^z\right]
\end{eqnarray}
with
\begin{equation}
n_i = \sum_l C_{li} \quad , \quad m_i = \sum_l C_{li}^2 \quad , \quad
a_{ij}=\sum_l C_{li} C_{lj} \quad , \quad b_{ij} = \sum_l C_{li}^2
C_{lj}
\end{equation}
Two-sites operators $\hat{W}_{ij}'$ do not commute when they share a
variable, and $\hat W'$ cannot be factorized as in the $K=1$ case. 
Yet, a cluster expansion in powers of $\alpha$ can be implemented. 
A detailed presentation of this cluster
expansion in a closely related context has been given
elsewhere~\cite{clusters}; the reader is referred to this
previous work for more details.
The method is presented below for a generic value of $K$. We call 
cluster and denote $F_r$ a maximal set of variables connected by clauses. 
Any formula $F$ can be decomposed as a conjunction of these clusters (note
that despite the similarities in denominations, these clusters have
nothing to do with the clustering phenomenon in configuration space).
Consider now a quantity $Q$, depending of the realization of
the disorder, namely of formula $F$, and additive with respect to the
clusters decomposition, $Q(F)=\sum_{F_r} Q(F_r)$. Calculation of  the
average $[Q]$ over the random ensemble formulas of $Q$ can be done
as follows. First, consider the different possible topologies $s$ of 
the clusters, compute $[Q]_s$, the quantity averaged over the choices 
of signs for a given topology {\em i.e.} whether variables appear
negated or not in a clause. Secondly, add up these contributions
with combinatorial factors giving the frequency of appearance of the
different topologies,
\begin{equation}
\frac{1}{M} [Q] = \sum_s \frac{1}{\alpha L_s} P_s [Q]_s
\label{generic_cluster_sum}
\end{equation}
The sum runs over the different topologies, $L_s$ is the number of sites
in such a cluster, and $P_s$ is the probability of a given site
belonging to an $s$-type cluster. In the thermodynamic limit, the
finite-size clusters which contribute to this sum are tree-like: 
$P_s = (\alpha K!)^{m_s} e^{-L_s \alpha K} K_s$, where 
$m_s = (L_s - 1)/(K-1)$ is the number of clauses of the cluster, 
and $K_s$ a symmetry factor. The smallest tree-like clusters are shown in
Figure~\ref{fig_clus}. Clauses are
represented by dashed lines (for $K=2$) or stars (for $K \ge 3$). In
principle, such an expansion is meaningful only below the
percolation threshold of the underlying random hypergraph, $\alpha_p =
1/(K(K-1))$. Beyond $\alpha_p$, a giant component appear,
whereas the series~(\ref{generic_cluster_sum}) takes into account
clusters of size $O(\ln N)$ only. However, rearranging the series as an
expansion in powers of $\alpha$ (expanding out the exponential in
$P_s$) allows to extend its domain of validity beyond $\alpha_p$, at
least for quantities that are not singular at the percolation
threshold.

This method can be employed here: the evolution operator $\hat{W}'$
is a sum of operators for each cluster $\hat{W}'(F_r)$. Operators 
attached to two distinct clusters commute with each other as they do
not have common variables. The energy operator $\hat{E}$ can also be
written as a sum over the clusters, thus the energy averaged over the
history of the algorithm has the property of additivity over clusters
if the time evolution is studied in terms of $u$. After averaging over
the formulas, the fraction of unsatisfied clauses reads,
\begin{equation}
\varphi_0(u) = \sum_s \frac{1}{\alpha L_s} P_s [\langle E
\rangle(u)]_s
\end{equation}
The evolution and energy operators for a cluster with $n$ variables are $2^n
\times 2^n$ matrices, $\hat{W}'_s$ and $\hat{E}_s$ respectively. With the 
help of a symbolic computation software one can easily study the small
finite-size clusters by computing
\begin{equation}
\left[\langle E \rangle (u)\right]_s = \left[ \langle O|\hat{E}_s
e^{\hat{W}'_s u}|{\bf S}(0)\rangle \right]_s
\end{equation}
where $\langle O|$ and $|{\bf S}(0)\rangle$ are now $2^n$ row and
column vectors, and the average is over the choices of negating or not
the variables in the clauses.

\begin{center}
\begin{figure}
\epsfig{file=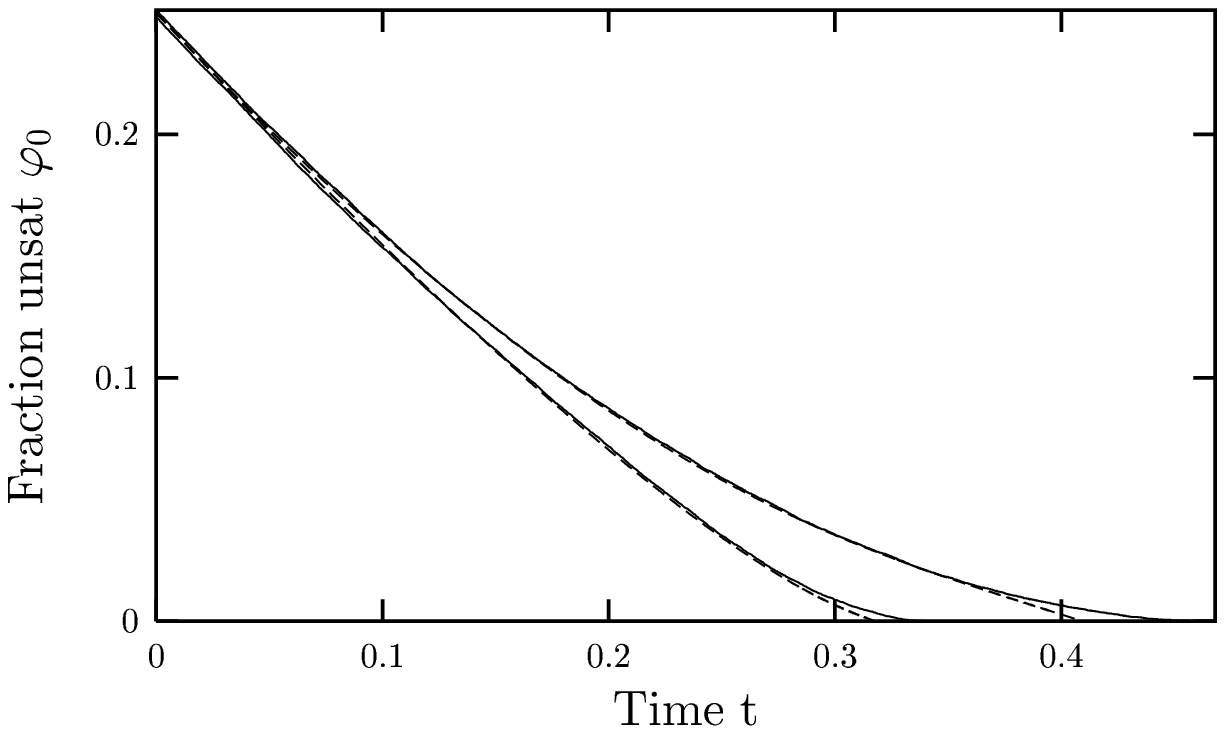,height=4cm}
\vskip .5cm
\caption{Fraction of unsat clauses as a function of time for $K=2$, 
$\alpha=0.6$ (top) and $\alpha=0.3$ (bottom). Dashed lines: 
cluster expansion prediction, solid lines (almost superimposed): numerical results 
(averaged over 10 runs for $N=10,000$).}
\label{eneK2}
\end{figure}
\end{center}

We have performed this task for clusters with up to four sites, their
contributions are summarized in Table~\ref{table_K2}. Up to order
$\alpha^2$, the expansion leads to
\begin{eqnarray} 
\varphi_0(u) &=& \frac{e^{-u}}{4}+\alpha\left(\frac{1}{4}e^{-u/2}
-\frac{3}{8}e^{-u}+ \frac{1}{8} e^{-2u}\right) \nonumber \\
&&+\alpha^2 \left(-\frac{3}{8} e^{-u/2} +\frac{19}{64} e^{-u} +
\frac{1}{8} e^{-3u/2}-\frac{9}{32}e^{-2u}+\frac{3}{64} e^{-3u} \right)
\nonumber \\ &&+\alpha^2 \left( \frac{\sqrt{3}-1}{32}
e^{-\frac{3+\sqrt{3}}{2}u} -\frac{\sqrt{3}+1}{32}
e^{-\frac{3-\sqrt{3}}{2}u} +\frac{2-\sqrt{2}}{16}
e^{-\frac{2+\sqrt{2}}{2}u} +\frac{2+\sqrt{2}}{16}
e^{-\frac{2-\sqrt{2}}{2}u}\right) + O(\alpha^3)
\label{phideu}
\end{eqnarray}
The typical value $\varphi_0(t)$ of the fraction of
unsatisfied clauses after $T=t\, M$ steps of the algorithm is
obtained from (\ref{phideu}) through the rescaling of time defined in
eqn. (\ref{eq_rescaling}). Our theory is compared 
in Figure~\ref{eneK2} to
numerical simulations. The agreement is excellent at the beginning of the
time evolution, and worsens at the end. A factor of explanation is
that during the last steps of the algorithm, the fraction of
unsatisfied clauses is low and thus for finite-size samples, the self
averaging hypothesis is violated.

Our approach is applicable to any value of $K$
(note however that the size of the matrices to be diagonalized
grows faster with the number of clauses in the clusters studied,
that is, with the order in $\alpha$ in the expansion) but is
restricted to the $\alpha <
\alpha_d$ regime. The finite-size tree clusters considered 
at any (finite) order in the expansion are indeed solved in linear
time by RandomWalkSAT. In Section~\ref{annealed_sec}, another
kind of approximation will allow to study the $\alpha > \alpha _d$ 
regime .
\begin{table}
\vspace{-.5cm}
\begin{center}
$$\begin{array}{c c c c c}
\hline
\multicolumn{1}{c}{\mbox{Type}} & 
\multicolumn{1}{c}{L_s} & 
\multicolumn{1}{c}{K_s} & 
\multicolumn{1}{c}{[\langle E \rangle (u)]_s} & 
\multicolumn{1}{c}{[t_{res}]_s} 
\\ 
\hline
a & 2  & 1 & \frac{1}{4}e^{-u} & 1/4
\\
b & 3 & 3/2 & \frac{1}{16}e^{-2 u} + \frac{5}{16}e^{- u} + \frac{1}{8}e^{-u/2}& 19/32
\\
c & 4 & 2 & \frac{1}{128}\left[ e^{-3 u} + 10 e^{-2u} + 4 e^{-3 u /2} + 53 e^{-u} + 20 e^{-u/2} + 2(2-\sqrt{2}) e^{-\frac{2+\sqrt{2}}{2}u} + 2(2+\sqrt{2}) e^{-\frac{2-\sqrt{2}}{2}u} \right]  & 125/128
\\
d & 4 & 2/3 & \frac{3}{256}\left[e^{-3 u} + 10 e^{-2u} + 25 e^{-u} + 32 e^{-u/2} - 2(1+\sqrt{3}) e^{-\frac{3-\sqrt{3}}{2}u} - 2(1-\sqrt{3}) e^{-\frac{3+\sqrt{3}}{2}u} \right]  & 259/256
\\
\hline 
\end{array}$$
\vspace{.25cm}
\caption{Contributions to the cluster expansions for $K=2$. See 
left panel of Figure~\ref{fig_clus}.}
\label{table_K2}
\end{center}
\end{table}

\subsection{Cluster analysis of resolution times}

As the number of steps needed to solve a formula is an additive 
quantity over clusters, it can be calculated along the lines exposed 
above for the expansion of $\varphi _0$.
Let us give two examples about how to compute the average time needed to
solve a cluster of fixed topology. 

The simplest example is a cluster made of a single 2-SAT clause,
$(x_1 \vee x_2)$. With probability $3/4$,
the initial random condition is already a solution. If not, it will
take one time step to solve the cluster, as any of the two possible
spin flips will lead to a solution. Thus the average
resolution time is $1/4$. Similar analysis can be
done on bigger clusters, even if the counting becomes increasingly
tiresome. As a byproduct, some flaws of the
RandomWalkSAT heuristic appear. 

\begin{center}
\begin{figure}
\epsfig{file=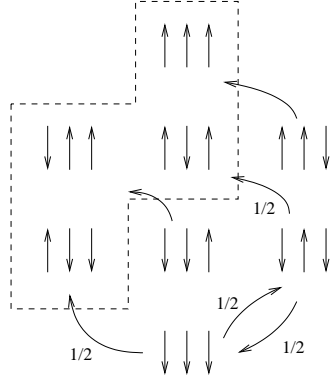,height=5cm} 
\vskip .1cm
\caption{An example of the behaviour of RandomWalkSAT on a finite cluster $(x_1 \vee x_2) \wedge (\bar{x}_2 \vee x_3)$. Blocks of spins represent configurations of the boolean variables, those in the shaded box are solutions. Arrows stand for possible transitions of the algorithm, see the text for details.}
\label{ex_time}
\end{figure}
\end{center}

The second example is the cluster
$(x_1 \vee x_2) \wedge (\bar{x}_2 \vee x_3)$. The eight
configurations of the variables are represented in
Figure~\ref{ex_time} with up or down spins corresponding to true or
false boolean variables. The four solutions are drawn in the shaded box.
Two configurations are turned into solutions in one
flip (single outcoming arrow). Flip of a spin 
from each of the remaining two configurations
(with two departing arrows), leads to a solution, or to the
other configuration. 
Thus the average resolution time starting from one of these
configurations is two timesteps. Finally, the average resolution time
for the cluster is $3/4$. To obtain the average over the signs of
this resolution time, one still has to make the same study for the
case where the two clauses are not contradictory on the central spin,
hence the value $19/32$ on line $b$ of Table~\ref{table_K2}.

Obviously, RandomWalkSAT can make ``bad'' choices on this very simple
example, and ``oscillate'' a few time steps between the two
configurations before finding a solution. One can imagine many local
search heuristics which would do better than RandomWalkSAT on the
example shown before. For instance, one could modify the algorithm so
that once an unsatisfied clause has been chosen randomly, it flips
preferably a variable with a low number of neighboors, or one with the
lowest number of contradicting clauses on it. The average resolution
time of any of these heuristics, as long as the information used to
choose the variable to be flipped remains local, can be studied by
such a cluster expansion. These simple enumerations could then
provide an useful test ground for new heuristics.

\begin{center}
\begin{figure}
\epsfig{file=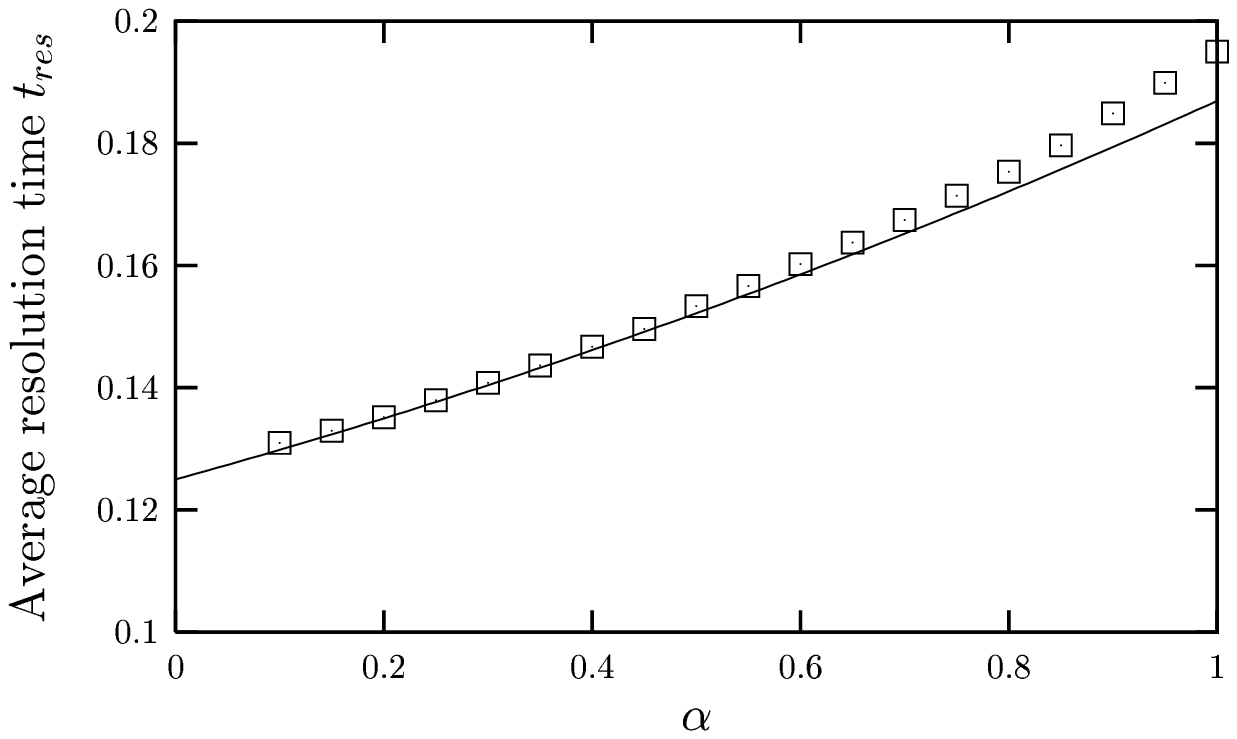,width=7cm} 
\caption{Average resolution time $t_{res}(\alpha , 3)$ for 3-SAT. 
Symbols: numerical simulations, averaged over $1,000$ runs for
$N=10,000$.  Solid line: prediction from the cluster expansion
(\ref{dev_cluster_tresK}).}
\label{K3_time_res}
\end{figure}
\end{center}

For general $K$, the enumeration of the possible histories of clusters
with three or less clauses leads to the quantities given in
Table~\ref{table_timeKg}, 
\begin{equation}
t_{res} (\alpha , K)= \frac{1}{2^K} + \frac{K(K+1)}{K-1}
\frac{1}{2^{2K+1}} \, \alpha + \frac{4K^6 + K^5 +6 K^3 -10 K^2 + 2
K}{3(K-1)(2K-1)(K^2-2)} \frac{1}{2^{3K+1}} \, \alpha^2 + O(\alpha^3)
\label{dev_cluster_tresK}
\end{equation}
This prediction is in good agreement with numerical simulations, see
Figure~\ref{K3_time_res} for results in the $K=3$ case.

The validity of expression (\ref{dev_cluster_tresK}) can be easily 
checked for $K=2$ from the findings of Section~\ref{sub_K2}.
Indeed, in terms of the time $t=T/M$, the fraction of unsat
clauses $\varphi_0(t)$ vanishes after a finite time $t_{res}$ given by
\begin{equation}
t_{res}=\lim_{u \to \infty} t(u) = \int_0^{\infty} du' \varphi_0(u')
\label{integral_tres}
\end{equation}
Integration of (\ref{phideu}) coincides  with prediction
(\ref{dev_cluster_tresK}) for $t_{res}(\alpha,2)$. 

\begin{center}
\begin{table}
$$\begin{array}{c c c c}
\hline
\multicolumn{1}{c}{\mbox{Type}} & L_s &  K_s (K!)^{m_s} / L_s & [t_{res}]_s \\
\hline a' & K & 1 & \frac{1}{2^K} \\ b' & 2K-1 &
\frac{K^2}{2} & \frac{1}{2^{2K}}\left[2^{K+1} + \frac{K+1}{K(K-1)}
\right] \\ c' & 3K-2 & \frac{K^3 (K-1)}{2} &
\frac{1}{2^{3K}}\left[3 \cdot 2^{2K} + 2^{K+1} \frac{K+1}{K(K-1)} +
\frac{4 K^4 + 9 K^3 +9 K^2 + 6K -4}{3 K^2 (K-1) (2K -1)(K^2 -2)}
\right] \\ d' & 3K-2 &  \frac{K^3}{6} & \frac{1}{2^{3K}}\left[3
\cdot 2^{2K} + 2^K \frac{3(K+1)}{K(K-1)} -
\frac{2K+1}{K^2(K-1)}\right]  \\
\hline
\end{array}$$
\caption{Contributions to the cluster expansion for the resolution time, 
general $K$. See right panel of Figure~\ref{fig_clus}.}
\label{table_timeKg}
\end{table}
\end{center}

\section{Approximate analysis of RandomWalkSAT}
\label{annealed_sec}
In this section, an analysis of RandomWalkSAT, based on a Markovian
approximation for the evolution equations (\ref{evolu1}) is
proposed. It allows a quantitative description of the main features
of RandomWalkSAT, namely the asymptotic energy and the (exponentially
small) probability of resolution in linear time 
for generic values of $K$ and $\alpha (>\alpha _d)$. 
The approximation scheme is also applied to the analysis of RandomWalkSAT 
on the XORSAT problem\cite{Crei}.

\subsection{Projected evolution}

Consider an instance of the $K$-SAT problem.  RandomWalkSAT defines a
Markov process on the space of configurations of the Boolean
variables, see eqn.~(\ref{evolu1}), of cardinality $2^N$. Call ${\bf
S}(T)$ the configuration of the Boolean variables at a given instant $T$
(number of flips) of
the evolution of the algorithm. An observable ${\cal R}$ is a function
of the configuration e.g. the number of clauses violated by ${\bf S}$.
The principle of the approach developed now is to
track the evolution of ${\cal R}(T) \equiv {\cal R}( {\bf S }(T))$, 
that is, of one number (or a low-dimensional vector) instead of the whole 
configuration of spins.  
To do so, we make use of the projection operator formalism exposed below.

Let us partition the configuration space into equivalence classes of
microscopic configurations ${\bf S}$ associated to the same value of
the macroscopic observable ${\cal R}({\bf S})$. We call $\Omega(
R)= \{{\bf S} | {\cal R}({\bf S})= R\}$ these classes, and
$|\Omega(R)|$ their cardinalities (number of configurations in
these classes).  Let us define the projection operator ${\cal
\hat{P}}$ through its entries,
\begin{equation}
\langle {\bf S}_1 | {\cal \hat{P}} | {\bf S}_2 \rangle 
= \frac 1{|\Omega( {\cal R}({\bf S}_1))|} \;\delta( {\cal R}
({\bf S}_1) - {\cal R}({\bf S}_2)) \quad ,
\end{equation}
where $\delta$ denotes the (vectorial) Kronecker function.  One can
easily check that it is indeed a projector, ${\cal \hat{P}}^2 = {\cal
\hat{P}}$, which connects only configurations within the same class.

Now consider the state vector $| {\bf S} (T) \rangle$ (\ref{sv}) and
its projection $|{\bf P} (T) \rangle \equiv {\cal \hat{P}} |{\bf S}
(T) \rangle $. Its components have the same value in each class, which
is the average of $| {\bf S} (T) \rangle$ over the microscopic
configurations in the class. Call $|{\bf Q} (T) \rangle = (1-{\cal
\hat{P}}) |{\bf S} (T) \rangle = |{\bf S} (T) \rangle- |{\bf P} (T)
\rangle$. From the master equation (\ref{evolu1}), we obtain
\begin{eqnarray}
|{\bf P} (T+1) \rangle &=& {\cal \hat{P}} \cdot \hat{W}_d |{\bf P} (T) \rangle 
 + {\cal \hat{P}} \cdot \hat{W}_d |{\bf Q} (T) \rangle  \nonumber \\  
|{\bf Q} (T+1) \rangle &=& (1-{\cal \hat{P}}) \cdot \hat{W}_d 
|{\bf P} (T) \rangle + (1-{\cal \hat{P}}) \cdot \hat{W}_d |{\bf Q} (T) \rangle
\end{eqnarray}
The second equation can be formally ``integrated'' by iteration, 
\begin{equation}
|{\bf Q} (T) \rangle =\sum_{T'=1}^T ((1-{\cal \hat{P}}) \cdot \hat{W}_d)^{T'} \;
|{\bf P} (T-T') \rangle 
\end{equation}
where the initial state vector $|{\bf S} (0)\rangle$  has been assumed 
to be uniform on each class, so that $|{\bf Q} (0) \rangle =0$.
Finally,
\begin{equation}
|{\bf P} (T+1) \rangle  = \sum_{T' =0}^T {\cal \hat{P}} \cdot 
\hat{W}_d \cdot ((1-{\cal \hat{P}}) \cdot \hat{W}_d)^{T'} |{\bf P} (T-T') \rangle 
\label{sum_proj}
\end{equation}
Equation (\ref{sum_proj}) expresses that, once coarse-gained by the
action of the projection operation, the dynamics is not Markovian any
longer. The principle of our approximation is precisely to omit all
memory effects, by neglecting non Markovian terms {\em i.e.} $T' \ge
1$ contributions in (\ref{sum_proj}), and averaging over disorder at
each time step,
\begin{equation}
|{\bf P} (T+1) \rangle  \simeq [ {\cal \hat{P}} \cdot 
\hat{W}_d ]\,  |{\bf P} (T) \rangle 
\label{sum_projbis}
\end{equation}
Obviously, the quality of
the approximation depends on the observable ${\cal R}$. We shall see
two examples in what follows.

\subsection{Transition matrix for the number of unsatisfied clauses}
\label{sub_annealB}
A natural choice we study in this and the next two subsections for the 
observable ${\cal R}$ is ${\cal M}_0$, which measures the numbers $M_0$ of 
clauses unsatisfied by each configuration of the variables. Defining the
bra
\begin{equation}
\langle M_0 | =  \sum _{{\bf S} \in \Omega (M_0)} \langle {\bf S}| 
\quad ,
\end{equation}
and the probability $\mbox{Prob}[M_0,T]=\langle {M_0} |{\bf P} (T)\rangle$
that the configuration of the variables is in class ${M_0}$ at time $T$,
we obtain within the Markovian approximation (\ref{sum_projbis}), 
\begin{equation} \label{evolumarkov}
\mbox{Prob}[{M_0'},T+1] = \sum_{M_0} A_{ M_0'\, M_0}\,
\mbox{Prob}[{M_0},T]\quad ,
\end{equation}
with $A_{ M_0' \,M_0}\equiv N_{ M_0'\,M_0} / D_{M_0}$ and
\begin{eqnarray}
N_{M_0' \, M_0}&=& \sum_{j=1}^N \sum_{{\bf S}} \delta \big( M_0 
-{\cal M}_0({\bf S}) \big)\, \delta\big({M_0'}-{\cal M}_0({\bf S}^j) \big) 
\, p_j({\bf S}) \nonumber \\
D_{M_0} &=& \sum_{{\bf S}}
\delta\big ({M}_0-{\cal M}_0({\bf S})\big)
\end{eqnarray}
where $p_j({\bf S})$ is the probability of flipping spin $k$ when
the system is in configuration ${\bf S}$ {\em i.e.} the number of unsat
clauses in which spin $j$ appears divided by the number of unsat
clauses. ${\bf S}^j$ denotes the configuration obtained from ${\bf S}$ by
flipping spin $j$. 
The meaning of our Markovian approximation is clear: the transition rate
from one value of the observable $M_0$ to another is the
average of the microscopic transition rates from one microscopic
configuration belonging to the first subset 
$\Omega (M_0)$  to another belonging to the
second one, with a flat average on the starting subset. At time $T$, the
only available information in the projected process 
is that the system is somewhere in the subset, 
and none of the corresponding microscopic configurations can be 
privileged.

To perform the average over the disorder, {\em i.e.} on the
random distribution of formulas and compute
$[A_{M_0' \,M_0}]=[N_{M_0'\, M_0}/D_{M_0}]$, 
we shall do the further approximation that the 
numerator and the denominator can be averaged separately.
This `annealed' hypothesis can be justified in some cases, see Section
\ref{secxorsat}.  After some combinatorics, we find
\begin{eqnarray} \label{evolusat}
[A_{M_0'\, M_0}] &=& \sum_{Z_u , Z_s} \frac{Z_u N}{K M_0}
{M_0 \choose Z_u} \,  {M-M_0 \choose Z_s} \,\left(1 -\frac{K}{N}\right)^
{M_0-Z_u} 
\left(\frac{K}{N}\right)^{Z_u} \times \nonumber \\
&& \left(1 -\frac{K}{(2^K - 1)\,N}\right)^{M-M_0-Z_s}
\left(\frac{K}{(2^K - 1)\,N}\right)^{Z_s} \, 
\delta (M_0'-M_0 + Z_u - Z_s) \ .
\end{eqnarray}
$Z_u$ is the number of unsatisfied clauses which contains the variable
to be flipped. All these clauses will become satisfied after the flip.
The factor $Z_u/(K M_0)$ represents the probability of 
flip of the variable, the factor $N$ coming from the sum over its index $j$. 
$Z_s$ is the number of clauses satisfied prior to the flip and 
violated after. The meaning of the Binomial laws is transparent.
Assume that the configuration violates $M_0$ clauses. 
In the absence of further information, the variable which 
is going to flip has probability $K/N$ to be present in
a given clause (there are ${N \choose K}$ possible $K$-uplets over
$N$ variables, and ${N-1 \choose K-1}$ which include a given variable). 
Furthermore, a satisfied clauses that contains the flipped variable
has a probability $1/(2^K-1)$ to become unsatisfied later.
$Z_u$ (respectively $Z_s$) is thus drawn from a binomial law with 
parameter $K/N$ (resp. $K/(N(2^K-1))$) , over $M_0$ (resp. $M-M_0$) 
tries. This reasoning unveils the physical significance of our Markovian 
approximation; we neglect all correlations between flipped variables 
and clauses that inevitably arise as the algorithm runs beyond the 
description in terms of the macroscopic variable $M_0$. 

\subsection{Average evolution and the metastable plateau}
\label{sub_annealC}
The evolution equation for the average fraction $\varphi _0 (T) = \sum_{M_0} 
M_0 \, \mbox{Prob}[{M_0},T] / M$ of unsatisfied clauses at time $T=t\,M$ is 
easily computed in the large size limit from 
(\ref{evolumarkov},\ref{evolusat}). In particular, the average fraction 
of unsat clauses equals
\begin{equation} \label{phi0si}
\varphi_0(\alpha,K,t) = 
\frac{1}{2^K}  + \frac{2^K - 1}{\alpha K 2^K} \left( e^{- \alpha K (1-2^{-K})^{-1} t} - 1 \right) \quad ,
\end{equation}
with $\varphi _0(0)=1/2^K$. Two regimes appear. If the 
ratio $\alpha$ is smaller than the critical value
\begin{equation} \label{alfd1}
\alpha_d(K)=\frac{2^K - 1}{K} \qquad ,
\end{equation}
the average fraction of unsat clauses $\varphi_0$ vanishes after a
finite time $t_{res}$. Typically, the algorithm will
find a solution after $t_{res}\times M$ steps (linear in $N$), and
then stops. Predictions for $\alpha_d$  are in good but not
perfect agreement with estimates from numerical simulations e.g.
$\alpha _d =7/3$ vs. $\alpha _d \simeq 2.7-2.8$ for 3-SAT.
The average resolution time $t_{res} (\alpha ,K)$ predicted within 
this approximation is given by the time at 
$\varphi_0 (t)$ in eqn (\ref{phi0si}) vanishes. It logarithmically 
diverges as $\alpha$ reaches the dynamical threshold at fixed $K$,
\begin{equation}
t_{res} (\alpha, K ) \sim - \frac{1}{2^K} 
\ln \big( \alpha_d (K) - \alpha \big) \ , \qquad 
\alpha \to \alpha_d (K)^- \quad .
\end{equation} 
On the contrary, when $\alpha >\alpha _d (K)$, 
$\varphi _0$ converges to a finite
and positive value 
\begin{equation}
\bar \varphi _0 (\alpha , K) =
\frac{1}{2^K} \; \left(1- \frac{\alpha_d(K)}{\alpha} \right)
\label{eq_recuit_plateau}
\end{equation} 
when $t\to\infty$ (Figure~\ref{plateau}). 
RandomWalkSAT is not able to find a solution and gets
trapped at a positive level of unsatisfied clauses. This situation arises
in the limit $T\propto N, N\to \infty$, and corresponds to the
metastable plateau identified in Section~\ref{sub_phen}.

\subsection{Large deviations and escape from the metastable plateau}
\label{sub_annealD}
As explained above, when $\alpha > \alpha _d$, the system gets almost
surely trapped in a metastable portion of the configuration 
space with a non zero number of unsatisfied clauses. 
Numerical experiments indicate the existence of an exponentially 
small-in-$N$ probability $\sim \exp ( N\, \bar \zeta (\alpha))$  
with $\bar \zeta < 0$
that this scenario is not correct, and that a solution is indeed found
in a linear time. We now make use of our Markovian hypothesis to
derive an approximate expression for $\bar \zeta$.

Contrary to the previous section, we now consider the large deviation
of the process with respect to its typical behaviour. This can be
accessed through the study of the large deviation function
$\pi({\varphi}_0,t)$ of the fraction
$\varphi_0$~\cite{Langer},
\begin{equation} \label{defpi}
\pi({\varphi}_0,t) = \lim _{N\to\infty} \frac 1N \, \ln
\mbox{Prob}\big[M_0= M \, {\varphi}_0 ,T = t \,M \big]
\end{equation}
Introduction of the reduced time is a consequence of the
following remark.
$O(1)$ changes in the fraction $\varphi _0$, that is, $O(N)$ changes
in the number $M_0$ of unsatisfied clauses are most likely to occur
after a number of flips of the order of $N$. 
To compute the large deviation function $\pi$, we introduce the 
generating function of $M_0$,
\begin{equation} \label{funcgen5}
G[{y},T] = \sum_{M_0} \mbox{Prob}[M_0,T] \; \exp ({y} \,
M_0 ) \quad ,
\end{equation}
where ${y}$ is a real-valued number. 
In the thermodynamic limit, $G$ is expected to scale exponentially with 
$N$ with a rate
\begin{equation}
g({y},t) \equiv  \lim _{N\to \infty} \frac 1N \, \ln 
G[ y, T = t\, M] = 
\max_{\varphi _0}[\pi({\varphi} _0,t) + \alpha
\, {y} \, {\varphi}_0] \ , \label{rt}
\end{equation}
equal to the Legendre transform of $\pi$ from insertion of definition
(\ref{defpi}) into eqn. (\ref{funcgen5}).
Using evolution equations (\ref{evolumarkov},\ref{evolusat}), we obtain the 
following equation for $g$,
\begin{equation} 
\frac{1}{\alpha} \frac{\partial g({y},t)}{\partial t} = -
y + \frac{\alpha K}{2^K -1} \big( e^y -1 \big) +
K \, \bigg( e^{-y} - 1 - \frac{1}{2^K - 1} (e^{y} -1 ) \bigg)
\, \frac{\partial g({y},t)}{\partial  y}   
\ .
\end{equation}
along with the initial condition 
\begin{equation}
g({y}, 0) = \alpha \, \ln \left(  1-\frac 1{2^K} +\frac {e^y}{2^K}  \right)
\quad .
\end{equation}
The average evolution studied in the previous section can be found again 
from the location of the maximum of $\pi$ or, equivalently, from the 
derivative of $g$ in ${ y}={ 0}$:
${\varphi}_0(t)=({1}/{\alpha})\, \partial g/\partial {y} 
({0},t)$.
The logarithm of the probability of
resolution (divided by $N$) at times large compared to $N$, but
very small compared to $\exp O( N)$, is given by
\begin{equation} \label{4df}
\bar \zeta (\alpha , K) = \pi({\varphi} _0 =0,t \to \infty ) 
= \int _0 ^{\tilde{y}(\alpha ) } dy \, z(y,\alpha) \ ,
\end{equation}
where $z (y,\alpha) = (y - \alpha K( e^y -1)/(2^K -1)) /
(K( e^{-y} - 1 - (e^{y} -1 )/(2^K -1)))$
and $\tilde{y}(\alpha)$ is the negative root of $z$.

\begin{center}
\begin{figure}
\epsfig{file=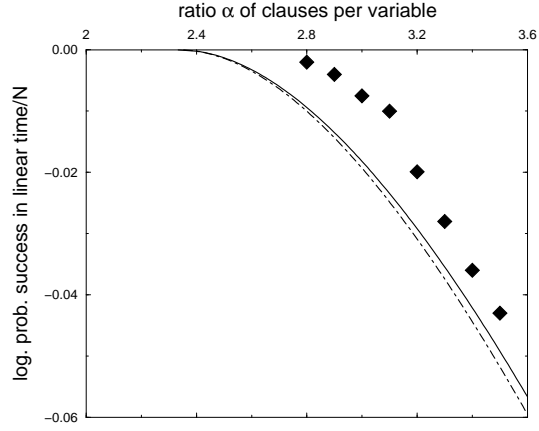,angle=-90,width=7cm}
\vskip .5cm
\caption{Large deviations for the 3-SAT problem. The logarithm 
$\bar \zeta $ of the probability 
of successful resolution (over the linear in $N$ time scale) is plotted
as a function of the ratio $\alpha$ of clauses per variables.
Predictions for $\bar \zeta (\alpha ,3)$ have been obtained within the
approximations of Section~\ref{sub_annealD} (equation (\ref{4df}), dot-dashed curve) and Section~\ref{sub_annealE} (fourth order solution of eqn (\ref{pde}), solid curve).  
Diamonds corresponds to (minus) the logarithm $\zeta$ of the average resolution times 
(averaged over 2,000 to 10,000 samples depending on the values of
$\alpha,N$, divided by $N$ and extrapolated to $N\to\infty$) obtained 
from numerical
simulations. Error bars are of the order of the size of the diamond symbol.
Sch\"oning's bound is $\bar \zeta \ge \ln(3/4) \simeq -0.288$.}
\label{prediczeta}
\end{figure}
\end{center}

Predictions for $\bar \zeta$ in the $K=3$ case are plotted 
in Figure~\ref{prediczeta}. They are compared to experimental
measures of $\zeta$, that is, the logarithm (divided by $N$) of the 
average resolution times. It is expected on intuitive grounds exposed in
Section~\ref{sub_phen} that $\zeta$ coincides with $-\bar\zeta$ 
(Figure~\ref{schema}). Despite the roughness of our Markovian approximation,
theoretical predictions are in qualitative agreement with numerical 
experiments.

\subsection{Taking into account clauses types}
\label{sub_annealE}
The calculation of Section IV.B can be extended to other observables 
${\cal R}$. In the following, we consider the case of a vectorial
observable $\vec {\cal M}$, with $K+1$ components. Our starting point is
the classification of clauses into `types'. 
A clause is said to be of type $i$, with $i = 0,\ldots ,K$, 
if the variables of the configuration ${\bf S}$ satisfy $i$
among $K$ of its literals. If $i=0$ the clause is unsatisfied while,
as soon as $i \ge 1$, the clause is satisfied.  Let us call $M_i({\bf
S})$ the number of clauses of type $i$, and $\vec {M}({\bf
S})=(M_0({\bf S}),\dots,M_K({\bf S}))$ the vector made with these
population sizes. Clearly, $\sum_i M_i({\bf S}) = M$ for any
configuration. If $M_0({\bf S})=0$ then ${\bf S}$ is a solution of the
formula.
Vector $\vec { M}$ is a natural characterization of the configuration of
variables (and of the instance), and contains essential information
about the operation of the algorithm. Indeed, the algorithm stops if the
number $M_0$ of unsatisfied clauses vanishes. In addition, at each
step of the algorithm, a single variable is flipped; clauses of type
$i$ become of type $i\pm 1$ if they include this variable, remain of
type $i$ otherwise. 

Within our Markovian annealed approximation, the probability 
$\mbox{Prob}[\vec { M},T]$ that the configuration of the variables is in 
class $\Omega (\vec { M})$ at time $T$ obeys the evolution equation,
\begin{equation} \label{evolumarkovm}
\mbox{Prob}[\vec {M'},T+1] = \sum_{\vec M} [A_{\vec M'\vec M}]
\;\mbox{Prob}[\vec { M},T]\quad ,
\end{equation}
with 
\begin{equation}
[A_{\vec M'\vec M}] = \sum_{\vec Z} \frac{N Z_0}{K
M_0}\; 
\delta({\vec M'}-{\vec M}-\Delta \cdot {\vec Z}) \; P({\vec Z}|{\vec M}) 
\end{equation}
where ${\vec Z}=(
Z_0,Z_1,Z_1^s,\dots,Z_i,Z_i^s,\dots,Z_{K-1},Z_{K-1}^s,Z_K)$ is a
$2K-1$ dimensional vector. Component $Z_i$ is the number of clauses 
of type $i$ where the variable which is going to flip appears. In $Z_i^s$
of these $Z_i$ clauses, this variable was one of the $i$ satisfying 
literals. It is not necessary to introduce components $Z_0^s$ and 
$Z_K^s$ for they have obvious values (respectively equal to $0$ and 
$Z_K$). $\Delta$ denotes a $(K+1) \times (2K-1)$ matrix such that 
$\Delta \cdot {\vec Z}$ gives the
change in the observable $\vec{\cal M}$ when the variable is flipped. The
$i$th line of $\Delta \cdot {\vec Z}$ reads $ -Z_i + Z_{i+1}^s +
(Z_{i-1}-Z_{i-1}^s)$. Clauses that contain 
the flipped variable and were of type $i$ prior to the flip are not any longer
of this type after the flip (hence the term $-Z_i$), those which
were of type $i+1$ and satisfied by the variable become of
type $i$ ($+ Z_{i+1}^s$), as those which
were of type $i-1$ and unsatisfied by the flipping variable 
($+Z_{i-1}- Z_{i-1}^s$).
The probability of ${\vec Z}$ conditioned to ${\vec M}$ is, as in
the simpler case of Section~\ref{sub_annealB} a product of Binomial laws,
\begin{equation} \label{pz}
P({\vec Z}|{\vec M})= \prod_{i=0}^{K} {M_i \choose
Z_i} \left(\frac{K}{N}\right)^{Z_i} \left(1-\frac{K}{N}\right)^{M_i-Z_i} 
\prod_{i=1}^{K-1}
{Z_i \choose Z_i^s} \left(\frac{i}{K}\right)^{Z_i^s}
\left(1-\frac{i}{K}\right)^{Z_i-Z_i^s} \ .
\end{equation}
Repeating the procedures of Sections~\ref{sub_annealC} and \ref{sub_annealD}, we find
\begin{itemize}
\item The average fraction of unsat clauses is calculated in Appendix~\ref{app_frac} and
reads
\begin{equation} \label{phi0}
\varphi_0(\alpha,K,t) = \frac{1}{2^K} + \frac{1}{\alpha K} \left[
\frac{1}{(1+\tanh(\alpha t))^K}-1 \right] \ .
\end{equation}
The critical ratio separating polynomial from exponential resolutions
is found at the same value as in Section~\ref{sub_annealB}, $\alpha_d(K)=(2^K-1)/K$. The average fraction of unsatisfied clauses on the plateau ($\varphi_0$ in the $t \to \infty$ limit) when $\alpha > \alpha _d (K)$ has also the same expression, cf eqn. (\ref{eq_recuit_plateau}). Note however that the finite time evolution differs in the two calculations.

When $\alpha<\alpha_d$, the resolution time $t_{res}$ is given by the 
vanishing of $\varphi_0$ in eqn (\ref{phi0}), and differs from its value
found within the simpler approximation of Section~~\ref{sub_annealB}.

\item The probability of easy (linear time) resolution is accessible
from the large deviation function
$\pi({\vec \varphi},t)$ of the fractions
$\varphi_i$ of clauses of type $0\le i\le K$. 
Its Legendre transform $g({\vec y},t)$ obeys the PDE
\begin{equation} \label{pde}
\frac{1}{\alpha} \frac{\partial g({\vec y},t)}{\partial t} = - y_0 +
y_1 + \sum _{i=0} ^K 
\big((K-i) e^{y_{i+1} -y_i} + i \,e^{y_{i-1} - y_i} -K\big) \, 
\frac{\partial g({\vec y},t)}{\partial  y_i}   
\end{equation}
along with the initial condition 
\begin{equation}
g({\vec y}, 0) = \alpha \, \ln \left( \sum _{i=0}^K \frac 1{2^K} {K \choose i}
 e^{y_i} \right)
\quad .
\end{equation}
The logarithm of the probability of
resolution (divided by $N$) at times large compared to $N$, but
very small compared to $\exp O( N)$, is given by
\begin{equation}
\bar \zeta (\alpha , K) = \max_{y_0} g(y_0,y_1=y_2=...=y_K=0,t)
\ .
\end{equation}
We have not been
able to caculate exactly $\bar\zeta$ for generic values of $\alpha$ and
$K$, but have resorted to a polynomial expansion of $g$ in powers of 
its arguments $y_i$. The expansion has been done up to order 4 with 
the help of a symbolic computation software for $K=3$, and up to order 
2 analytically for any $K$. Calculations are exposed in Appendix~\ref{app_PDE}.
 Predictions for $\bar \zeta (\alpha)$ in the $K=3$ case are plotted 
in Figure~\ref{prediczeta}. 
\end{itemize}

\subsection{The large $K$ limit}
\label{sub_annealF}
Comparison between results of Sections~\ref{sub_annealB} and \ref{sub_annealE} shows that
the output of the calculation quantitatively depends on the observable under
study. However, we may expect some simplification to take place
for large $K$. In this limit, if a clause gets unsatisfied twice, or more 
(but $\ll K$ times), it is very unlikely that each variable will be 
flipped more than once, and memory effects are lost. 
Therefore, the Markovian annealed approximation is expected to become 
correct. However, to avoid a trivial limit, the ratio $\alpha$ of 
clauses per variable must be rescaled accordingly. Inspection of the 
above result (\ref{alfd1}) indicate that the correct scaling is
$\alpha , K\to \infty$ at fixed ratio $\alpha ^* = \alpha\, K/ 2^K$.
The dynamical threshold separating linear from exponential resolutions
is located in
\begin{equation}
\alpha _d^* = 1 \quad .
\end{equation}
As the critical threshold of $K$-SAT is known to scale as 
$\alpha _c (K) \sim 2^K \, \ln 2$ for large $K$\cite{mz,achliperes}, 
instances are always satisfiable on the reduced $\alpha ^*$ scale.

For $\alpha ^* < 1$, the initial fraction of unsatisfied clauses is 
$\simeq 1/2^K$, and decreases by $O(1)$ per unit of reduced time $t$, 
giving $t_{res} \sim 1/2^K$. For the same reason, the height
$\varphi _0$, which is reached after a $O(1)$ relaxation time when 
$\alpha ^*>1$, is of the order of $1/2^K$. It is therefore natural to define
the rescaled fraction of unsatisfied clauses through
\begin{equation}
\varphi ^* _0 (\alpha ^*, t^*) = \lim _{K \to \infty}  
2^K \; \varphi _0 \left( \alpha ^* \,
2^K/K, K , t^* / 2^K  \right) \ , 
\end{equation}
from which we obtain the rescaled resolution time $t_{res}^* (\alpha ^*)$
for $\alpha ^*<1$ (vanishing of $\varphi _0^*$) and plateau
height $\bar \varphi _0^* (\alpha ^*)$ for $\alpha ^*> 1$ (limit value
of $\varphi _0^*$ at large rescaled times).
The two schemes of approximation of Section~\ref{sub_annealB} and \ref{sub_annealE} both yield
\begin{eqnarray}
\varphi _0 ^* (\alpha ^*,t^*) &=& 1+\frac 1{\alpha ^*} \left(
e^{-\alpha ^*\, t^*}-1 \right)
\\
t_{res} ^* (\alpha ^* ) &=& -\frac 1{
\alpha^*} \, \ln \big(1- \alpha^* \big)  = 
1 + \frac 12 \,  \alpha ^* + \frac 13 \,  (\alpha^*) ^2 + 
O\big((\alpha ^*)^3\big) \label{asymtres} \\
\bar{\varphi} _0 ^* (\alpha ^* ) &=& 1-\frac 1{
\alpha^*}  = \alpha^* - 1 + O\big((\alpha ^*-1)^2\big)  
\ ,
\end{eqnarray}
Note that the small $\alpha ^*$ expansion (\ref{asymtres}) for the 
resolution time coincides with the exact expansion obtained 
from eqn (\ref{dev_cluster_tresK}) with the above rescaling of $\alpha$ 
and $K$. We conjecture that the equality holds for higher orders
($\ge 3$) in $\alpha ^*$, and that the above expressions for 
$\varphi _0 ^*( \alpha ^*, t^*)$ and thus for 
$t^*_{res}(\alpha ^*), \bar \varphi _0^*(\alpha ^*)$ are correct.

The logarithm of the probability of linear time resolution
for $\alpha >\alpha _d(K)$ needs  to be rescaled too,  
\begin{equation} \label{resca}
\bar \zeta ^* (\alpha ^*) = \lim _{K \to \infty}  
K\,  \bar \zeta (\alpha ^* \, 2^K/K,K) 
\end{equation}
to acquire a well defined limit when $K\to \infty$. 
The scalar approximation of Section~\ref{sub_annealD} gives the
asymptotic result, 
\begin{equation} \label{resca89}
\bar \zeta ^* (\alpha ^*) = \int _0 ^{\tilde{y}(\alpha ^*)}
dy \, \frac{y-\alpha ^* (e^y -1)}{e^{-y}-1}
= - (\alpha^* - 1) ^2 + 
O\big((\alpha ^*-1)^3\big) \ , 
\end{equation}
where $\tilde{y} (\alpha ^*)$ is the negative root of the numerator in the 
above integral. The quadratic resolution of the PDE arising from the study of Section~\ref{sub_annealE} (cf. Appendix~\ref{app_PDE}) also leads to this result around $\alpha^*=1$.
Unfortunately, the exact results obtained in Section~\ref{sec_exact}
are of no help to confirm identity (\ref{resca89}).

\subsection{The XORSAT case}
\label{secxorsat}

XORSAT is a version of a satisfiability problem, much simpler than SAT
from a computational complexity point of view\cite{Crei,crei2,xorsat1,xorsat2}. 
One still draws $K$-uplets of variables, but each clause bears only one
`sign' (instead of one for each variable in the KSAT version), and the
clause is said to be satisfied if the exclusive OR (XOR) of its
boolean variables is equal to the `sign' of the clause. For a given
clause, there are $2^{K-1}$ satisfiable assignments of the variables,
and also $2^{K-1}$ unsatisfiable assignments, in deep contrast with
SAT where these numbers are respectively equal to $2^K-1$ and
$1$. XORSAT may be recast as linear algebra problem, where a set of
$M$ equations involving $N$ Boolean variables must be satisfied modulo
2, and is therefore solvable in polynomial time by
various methods e.g. Gaussian elimination. Nevertheless, it is
legitimate to ask what the performances of local search methods as
RandomWalkSAT are for this kind of computational problem.

\begin{center}
\begin{figure}
\epsfig{file=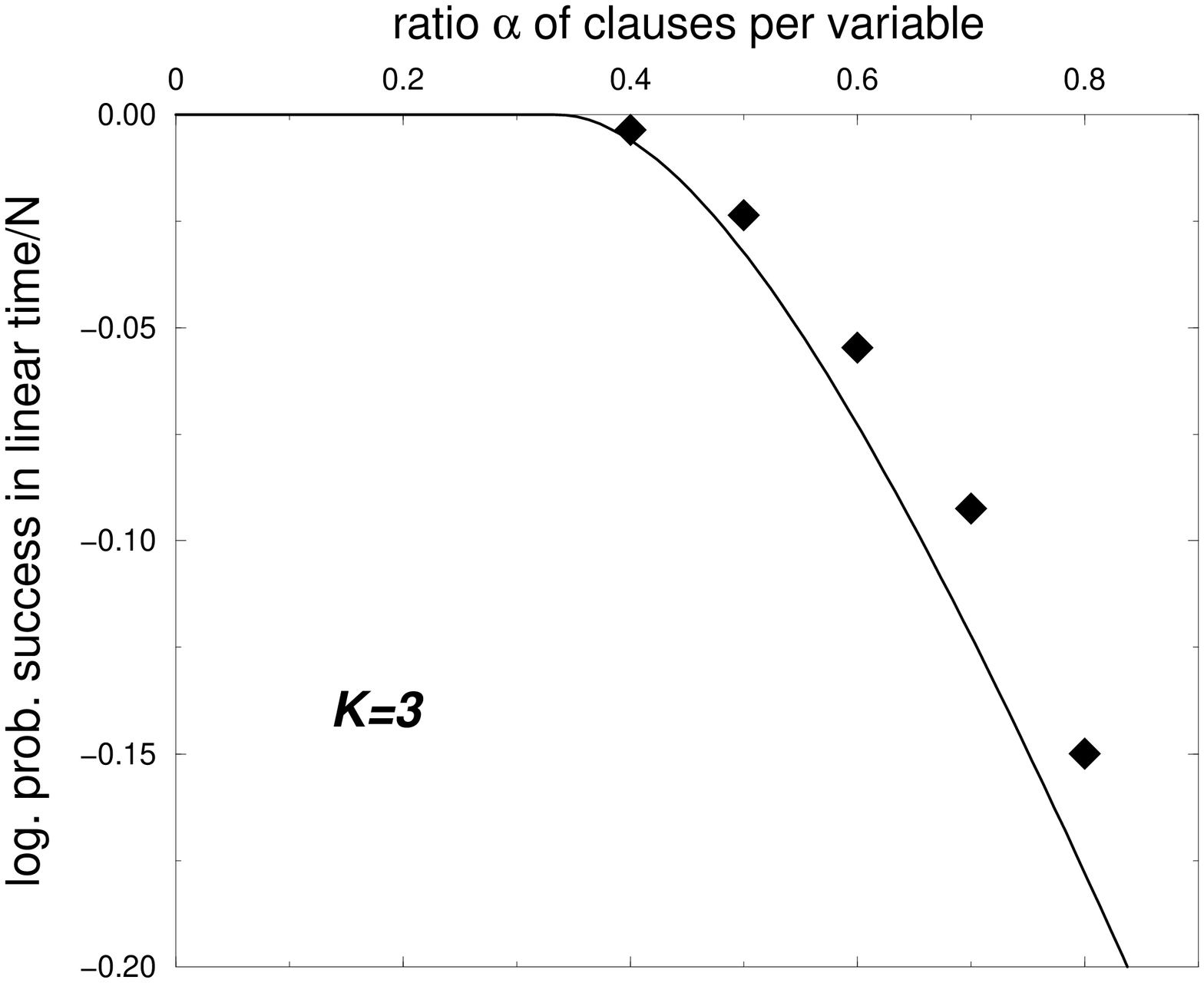,angle=-90,width=7cm}
\hskip .5cm
\epsfig{file=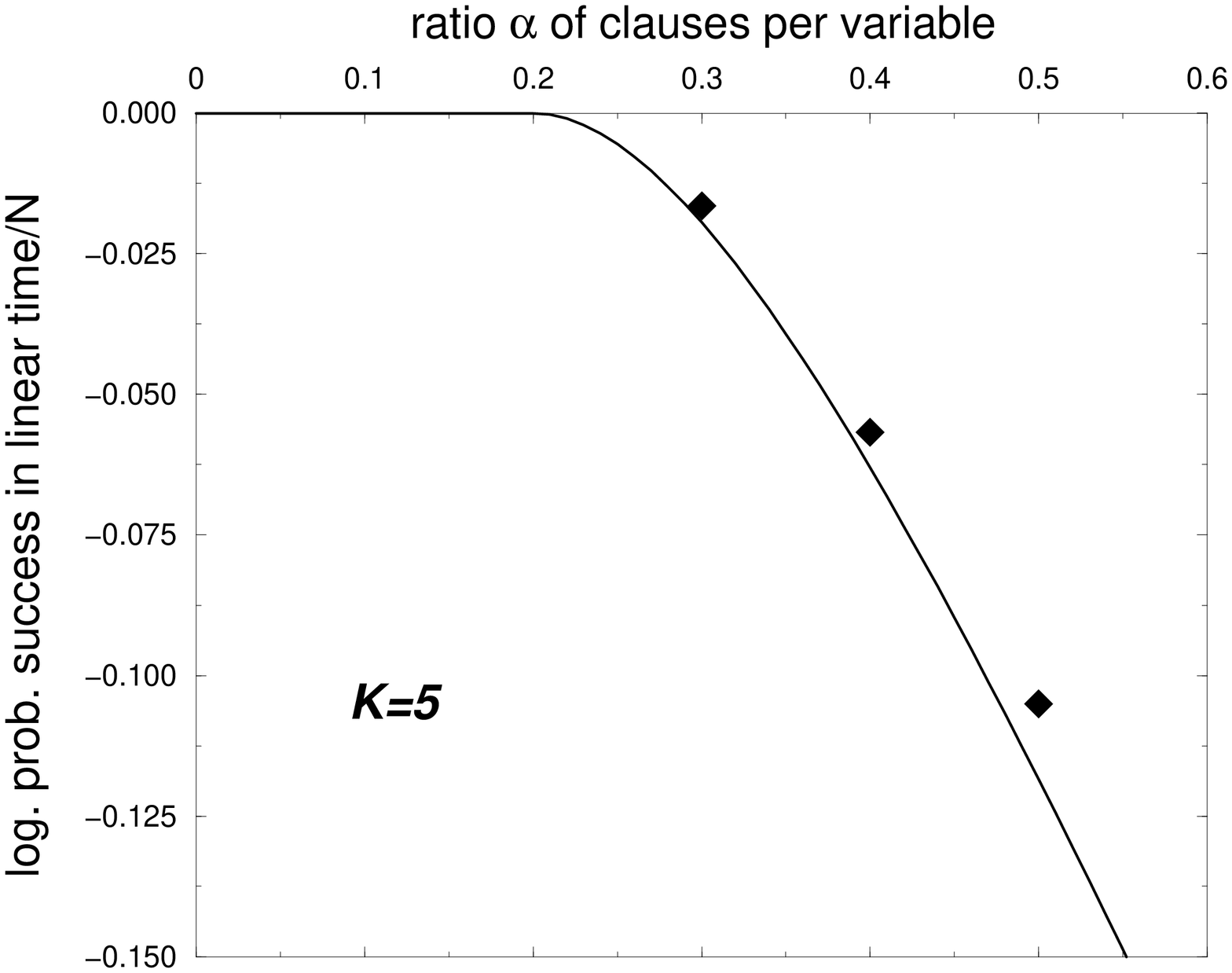,angle=-90,width=7cm}
\vskip .5cm
\caption{Large deviations for the K-XORSAT problem for $K=3$ (left) and
$K=5$ (right). The logarithm $\bar \zeta $ of the probability 
of successful resolution (over the linear time scale) is plotted
as a function of the ratio $\alpha$ of clauses per variables. Diamonds
corresponds to (minus) the logarithm $\zeta$ of the average resolution times 
(averaged over 10,000 samples, divided by $N$ and extrapolated to 
$N\to\infty$) obtained from numerical
simulations. Error bars are smaller than the size of the diamond symbol. }
\label{predicxor}
\end{figure}
\end{center}

A fundamental feature of XORSAT is that, whenever a spin is flipped,
all clauses where this spin appears change status: the satisfied ones
become unsatisfied, and {\em vice versa}.  There is thus no need to
distinguish between clauses satisfied by a different number of
literals, and the macroscopic observable we track is the number $M_0$ of
unsatisfied clauses for configuration ${\bf S}$ as in Section~\ref{sub_annealB}. It is an easy check that the transition matrix $[A]$ for XORSAT is 
given by eqn (\ref{evolusat}) where $2^K - 1$ is replaced with 1. 
Main results are:
\begin{itemize}
\item The average fraction of unsatisfied clauses $\varphi _0(t)$ reads
\begin{equation}
\varphi _0(\alpha, K, t) = 
\frac{1}{2} + \frac{1}{2 \alpha K} (e^{- 2 \alpha K t}-1)
\quad ,
\end{equation}
and becomes asymptotically strictly positive if the ratio $\alpha$ of
clauses per variables exceeds $\alpha_d (K)= 1/K$, smaller than the clustering
and critical ratios, $\alpha _s \simeq 0.818$ and $\alpha _c\simeq 0.918$ for
$K=3$ respectively\cite{xorsat1,xorsat2}. The overall picture
of the algorithm behavior is identical to the SAT case. 
\item When $\alpha > \alpha _d (K)$, the average fraction of 
unsatisfied clauses on the plateau is given by
\begin{equation}
\bar \varphi _0 (\alpha ,K) = \frac 1{2} \; \left(1- \frac 1
{\alpha\, K} \right) \ .
\end{equation} 
\item The partial differential equation for the generating function is,
\begin{equation}
\frac{1}{\alpha} \frac{\partial}{\partial t} g(y,t) = - y + \alpha K (e^y -
1)+  K (e^{-y} - e^{y}) \frac{\partial}{\partial y} g(y,t) 
\end{equation}
with $g(y,0)=\alpha \ln [(1+e^y)/2]$. Resolution in the large time limit
is straightforward, with the results shown in Figure~\ref{predicxor}.
The agreement with numerics is good, especially as $K$ grows. 
\end{itemize}

The relatively simple structure of XORSAT makes possible the test of 
some of the approximations done. We show in Appendix~\ref{app_XOR} that the annealed
hypothesis done in the calculation of the evolution matrix
$A$ is justified in the thermodynamic limit. The validity of this
approximation in the case of 3-SAT (Section~\ref{sub_annealB}) is not established for finite $K$. 

As for $K$-SAT, quantities of interest have a well defined large
$\alpha, K$ limit provided the ratio $\alpha^*=\alpha/\alpha _d(K) =
\alpha\, K$ is kept fixed;
\begin{eqnarray}
\varphi ^* _0 (\alpha ^*, t^*) &=& \lim _{K \to \infty}  
\varphi _0 \left( \alpha ^*/K, K ,  t^*\right) = 
\frac{1}{2} + \frac{1}{2 \alpha ^*} (e^{- 2 \alpha ^*\, t^*}-1) \\
\bar\zeta ^*  (\alpha ^*) &= & \lim _{K\to \infty} K\, \bar \zeta \left( 
\alpha ^* / K ,  K \right) =  \int _0 ^{\tilde{y}(\alpha ^* )} dy \;
z (y,\alpha ^*) \\
&=& -\frac 12 ( \alpha ^* -1) ^2 + O\big(( \alpha ^* -1) ^3 \big) \quad , 
\nonumber
\end{eqnarray}
where $z (y,\alpha ^*) = (y-\alpha ^* (e^y-1))/(e^{-y} - e^y)$
and $\tilde{y}(\alpha ^*)$ is the negative root of $z$.

\section{Conclusion and perspectives}
\label{sec_conclu}
In this paper, we have studied the dynamics of resolution of a simple
procedure for the satisfaction of Boolean constraints, the RandomWalkSAT
algorithm. We have shown using complementary techniques (expansions
and approximations) that, for randomly drawn input instances, RandomWalkSAT
may have two qualitatively distinct behaviours. Instances with small
ratios $\alpha$ of clauses per variable are almost surely solved in a
time growing linearly with their size. On the contrary, for ratios
above a threshold $\alpha _d$, the dynamics gets trapped for an
exponentially large time in a region of the configuration space with a
positive fraction of unsatisfied clauses. A solution is finally
reached through a large fluctuation from this metastable
state.

The freezing taking place at $\alpha _d$ does not seem to be
related to the onset of clustering between solutions\cite{giul}.
Indeed the value of $\alpha _d$ is expected to change with the
local search rules. It
would be interesting to pursue the study initiated in the present work
to understand if and how the existence of this dynamical threshold is
related to some property of the (static) energy landscape, as
in mean field models of spin glasses\cite{leti}.  Another
useful improvement would be to go beyond the Markovian approximation
of Section~\ref{annealed_sec}. Unfortunately, keeping a finite (with respect to $N$)
number of retarded terms in (\ref{sum_proj}) should not be sufficient
to achieve this goal.  Improvements will require to take into account
an extensive number of terms, or to extend the quantum formalism of
Section III to the study of the metastable phase. 
Another possible direction of research would be to use projection operators on observables of extensive dimension~\cite{and,dynamicalreplica}.
Our Markovian approximation, expected to be exact in the large
$K$ limit, should be a 
starting point for a systematic large $K$ expansion of the quantities
of interest (plateau height, life time of the metastable regime, ...). 
Finally, extension
of our analysis to more sophisticated local search heuristics would be
useful.
 
\begin{acknowledgments} 
We thank W. Barthel, A.K. Hartmann, M. Weigt for oral
communication of their results prior to publication\cite{and},
and S. Cocco, L. Cugliandolo 
and A. Montanari for useful discussion and comments. The quantum
formalism presented in Section III is inspired from a previous
unpublished work of one of us with G. Biroli, whom we are very
grateful to. We also thank C. Deroulers for his help in the 
perturbative resolution of eqn (\ref{pde}) (Section III.C).
The present work was partly supported by the French Ministry of
Research through the ACI Jeunes Chercheurs ``Algorithmes d'optimisation 
et syst{\`e}mes d{\'e}sordonn{\'e}s quantiques''.
\end{acknowledgments}

\appendix
\section{Generating function for the average evolution}
\label{app_frac}
 Defining ${\vec \varphi} = 
(\varphi_0,\dots,\varphi_K) =\sum_{\vec M} \mbox{Prob}[\vec M , T] \, \vec M /M $ and the reduced time $t=T/M$, one get
from eqn (\ref{evolumarkov}), 
\begin{equation}
\frac{d {\vec \varphi}}{dt} = {\vec v} + \alpha\, K \, {W}_r \cdot {\vec \varphi} \quad ,
\label{eq_phi}
\end{equation}
with ${\vec v}$ a $K+1$-dimensional vector and $W_r$ a $(K+1) 
\times (K+1)$ matrix defined as
\begin{equation}
{\vec v} = \left( \begin{array}{c} -1\\ 1 \\ 0 \\ \vdots \\0 \end{array}
\right) \ , \qquad {W}_r = \left( \begin{array}{c c c c c c c} -1 &
\frac 1K & 0 & 0 & 0 & \ldots & 0\\ 1 & -1 & \frac 2K & 0 & 0 & \ldots
& 0\\ 0 & \frac{K-1}K & -1 & \frac 3K & 0 & \ldots & 0 \\ \vdots &
\vdots & \vdots & \vdots & \vdots & \ldots & \vdots \\ 0 & 0 & 0 & 0 &
\ldots& \ldots & -1 \\
\end{array} \right) 
\end{equation}
To solve equation (\ref{eq_phi}), it results convenient to introduce
$\Phi(x,t)$, the polynomial in $x$ whose coefficients are the 
fractions $\varphi_i$ we want to determine,
\begin{equation}
\Phi(x,t)=\sum_{j=0}^K \varphi_j(t) x^j
\end{equation}
The set of $K+1$ linear coupled differential (\ref{eq_phi}) reduces to 
a partial differential equation on $\Phi(x,t)$ :
\begin{equation}
\frac{\partial \Phi}{\partial t} (x,t) = -1+x + \alpha K (x-1) 
\Phi(x,t) + \alpha (1-x^2) \frac{\partial \Phi}{\partial x} (x,t)
\label{eq_app}
\end{equation}
At initial time, the variables are chosen randomly to be true or false, 
without any correlation with the formula studied. Thus, the number of 
satisfied literals in a clause obeys a binomial law with parameter 
$1/2$: $\varphi_j(0) = \frac{1}{2^K} {K \choose j}$. 
The initial condition on $\Phi$ reads thus
\begin{equation} \label{ic}
\Phi(x,0) = \left( \frac{1+x}{2}\right)^K
\end{equation}
Setting $\Psi(x,t)=\Phi(x,t)+{1}/{(\alpha K)}$, the constant terms in 
eqn (\ref{eq_app}) can be eliminated, with the resulting PDE for $\Psi$,
\begin{equation}
\frac{\partial \Psi}{\partial t} (x,t) =  \alpha K (x-1) \Psi(x,t) + 
\alpha (1-x^2) \frac{\partial \Psi}{\partial x} (x,t)
\end{equation}
This can be transformed into a wave equation on 
$\chi(x,t)=(1+x)^{-K} \Psi(x,t)$ :
\begin{equation}
\frac{\partial \chi}{\partial t} (x,t) = \alpha (1-x^2) \frac{\partial 
\chi}{\partial x} (x,t)
\end{equation}
This equation is solved in terms of an arbitrary function of a 
single argument,
\begin{equation}
\chi (x,t)= \omega \left( \frac{1}{\alpha} \tanh^{-1}(x) + t\right)
\end{equation}
Knowledge of $\Phi(x,0)$ for all $x$ (\ref{ic}) allows us to determine 
unambiguously $\omega$,
\begin{equation}
\omega(u) = \frac{1}{2^K} + \frac{1}{\alpha K} [1+\tanh(\alpha u )]^{-K}
\end{equation}
Going backwards, we obtain the expression of the generating function
\begin{equation}
\Phi(x,t) = \left( \frac{1+x}{2} \right)^K + \frac{1}{\alpha K} 
\left\{ \left[ \frac{1+x \tanh(\alpha t)}{1+ \tanh(\alpha t)} 
\right]^K - 1 \right\}
\end{equation}
and of the fractions of clauses of type $i$ through an expansion of the
latter in powers of $x$,
\begin{equation}
\varphi_j(t) ={K \choose j}\left[ \frac{1}{2^K} + \frac{1}{\alpha K} \frac{(\tanh (\alpha t ))^j}{(1+\tanh (\alpha t ))^K} - \frac{\delta_{j 0}}{\alpha K} \right] \ .
\end{equation}

\section{Perturbative resolution of the large deviation PDE}
\label{app_PDE}
In this appendix we sketch the resolution of PDE~(\ref{pde}) in the
long time limit, where the function $g$ becomes independent of
time. We expand it in powers of its arguments:
\begin{equation}
g({\vec y},t \to \infty) = \sum_{i=0}^K a_i y_i + \frac{1}{2}
\sum_{i,j=0}^K a_{ij} y_i y_j + \dots
\end{equation} 
Plugging this expansion into eqn.~(\ref{pde}), one obtains by
identification of the monomials in $y_i$ an infinite set of linear
equations on the coefficients of $g$, which can be solved order by
order. Constraint $\sum_i \varphi_i =1$
imposes a condition on $g$,
\begin{equation}
g({\vec y} + c \, {\vec 1}) = \alpha \, c + g({\vec y})
\label{jauge}
\end{equation}
where ${\vec 1}=(1,1,\dots,1)$ and $c$ is an arbitrary constant.

In the case $K=3$, we have solved the set of equations on the
coefficients of $g$ up to order $4$ in the $y_i$s with the help of a symbolic
computation software. To calculate $\bar \zeta$  we need to know
$g$ as a function of $y_0$ only, with $y_i=0 \,
\forall i\ge 1$. We find
\begin{equation}
g(y_0)=\frac{3 \alpha - 7}{24} y_0 + \frac{105 \alpha - 94}{1920}
y_0^2 + \frac{26460 \alpha + 10753}{193560} y_0^3 + \frac{29645 \alpha
+66244}{18923520} y_0^4 + O(y_0^5)
\end{equation}
When $\alpha>\alpha_d(K=3)=7/3$, this function has a non trivial
extremum, in which $g$ takes the value $\bar{\zeta}(\alpha)$.

We now explain the resolution at quadratic order for a generic value
of $K$. At linear order, following the calculation 
exposed in Appendix~A,
\begin{equation}
a_i = \alpha \lim_{t \to \infty} \varphi_i(t) = \frac{1}{2^K} {K
\choose i} \left( \alpha + \frac{1}{K}\right) - \frac{\delta_{i0}}{K}
\end{equation}
In particular $a_0=(\alpha-\alpha_d(K))/2^K$.  Then considering the
monomials of second order in the expansion of the equation, a set of
$(K+1)(K+2)/2$ linear equations determine the coefficients
$a_{ij}$. As we shall not try to solve the equation at higher orders
for this generic case, we need only $a_{00}$. Again we introduce a
generating function to turn the discrete algebraic problem into an
analytic one. $f(s)\equiv \sum_{i,j} a_{ij} s^{i+j}$ obeys the ordinary
differential equation,
\begin{equation}
2 K f(s) - (s+1) f'(s) = (s-1) \left(1- (1 + \alpha
K)\left(\frac{1+s^2}{2}\right)^{K-1} \right)
\end{equation}
with the condition $f(1)=0$ stemming from~(\ref{jauge}). This equation
can be easily solved, yielding
\begin{equation}
a_{00}=f(0)=\frac{2^K - 1}{2^{2 K}} \left( \alpha - \frac{(K-1) 2^{2K}
+ 2^K + 2 K - K 2^{K+1}}{K (2K - 1) (2^K - 1)} \right)
\end{equation}
At this order of the expansion, the extremum of $g$ in the subspace
$y_i=0, \, \forall i \ge 1$ is reached in $y_0=-a_0/a_{00}$ and leads
to $\bar{\zeta}(\alpha,K)=-a_0^2/(2 a_{00})$.

\section{Validity of the annealed hypothesis for the XORSAT problem}
\label{app_XOR}
We justify in this appendix the annealed average of the Markovian
transition matrix in the XORSAT case. Our analysis is based on
Chebyshev inequality~\cite{Spencer}: a positive integer valued random
variable with a variance negligible with respects to the square of
its average is sharply peaked around its mean value. 
Call $D_U=\sum_{\bf S} \delta( U
-U({\bf S}))$ the number of configurations with $U$ unsatisfied
clauses. The first moment of $D_U$ over the distribution of XORSAT instances
is easy to compute. After averaging,
the $2^N$ configurations of the variables contribute equally to the
sum; for each of them, the number $U$ of unsatisfied clause has a
binomial distribution of parameter $1/2$ among the $M$ clauses. In the
thermodynamic limit, using Stirling's formula and denoting $\varphi_0
= U/M$,
\begin{equation}
[D_U] \sim e^{N f_1(\varphi_0,\alpha)} \quad , \quad
f_1(\varphi_0,\alpha)=\ln 2 + \alpha (-\varphi_0 \ln \varphi_0 -
(1-\varphi_0) \ln (1-\varphi_0) - \ln 2)
\end{equation}
up to polynomial corrections. Suppose that $\varphi_0<1/2$.
Call $\varphi_0^{(1)}(\alpha)$ the root of $f_1$ at fixed $\alpha$.  
It is a growing function of $\alpha$,
vanishing for $\alpha \le 1$, and monotonously increasing to $1/2$ as
$\alpha$ gets large.
For $\varphi_0 > \varphi_0^{(1)}(\alpha)$,
$f_1$ is positive, and $[D_U]$ exponentially large. 
When $\varphi_0 < \varphi_0^{(1)}(\alpha)$, $[D_U]$ is exponentially
small. 

\begin{center}
\begin{figure}
\epsfig{file=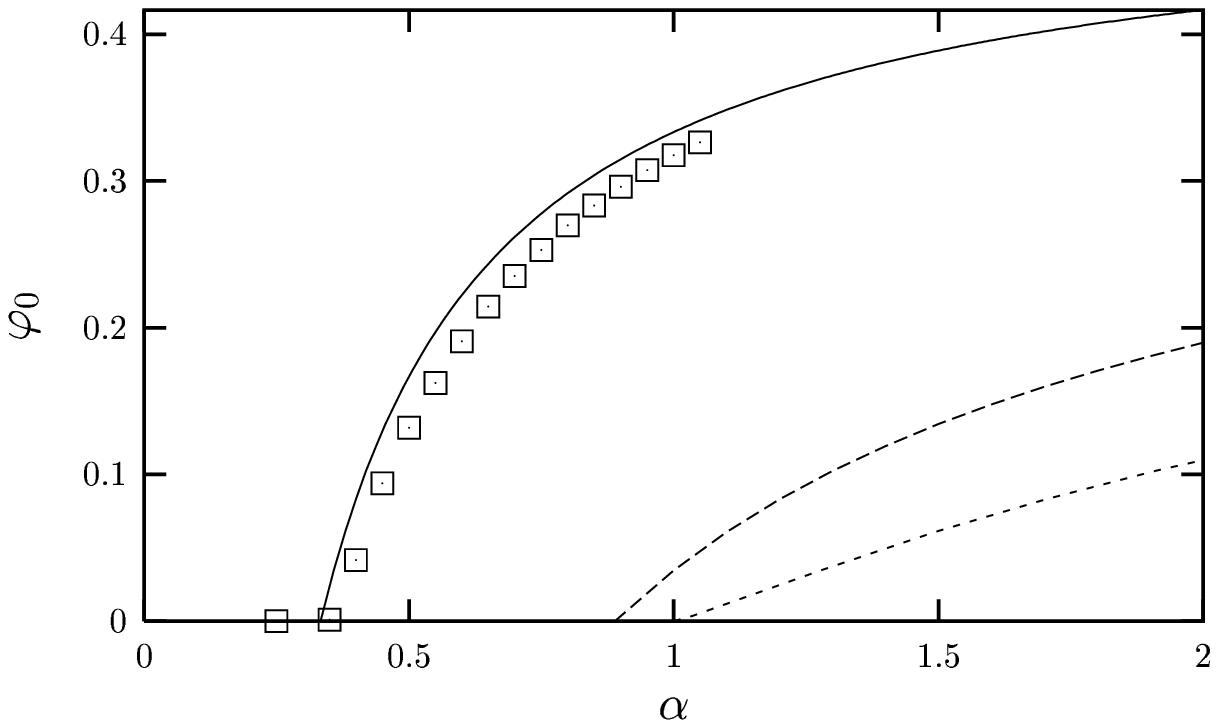,height=5cm} \vskip .5cm
\caption{Study of the moments of $D_U$ for 3-XORSAT. Solid line: Markovian
annealed prediction 
for the asymptotic fraction of unsat clauses $\varphi_0$. Long-dashed 
line: $\varphi_0^{(2)}$. Short-dashed line: $\varphi_0^{(1)}$. Symbols: asymptotic fraction of unsat clauses on the plateau, obtained through numerical simulations.}
\label{fig_moments}
\end{figure}
\end{center}

Consider now the second moment $[D_U^2]$ and its leading behaviour
$[D_U^2]\sim \exp[N f_2(\varphi_0,\alpha)]$. We introduce the generating
function
\begin{equation}
\sum_U [D_U^2]\, e^{-2 x U} = \int_0^{2 \pi} \frac{d\theta}{2 \pi}
\sum_{{\bf S}_1,{\bf S}_2} \left[ e^{-x(U({\bf S}_1)+U({\bf S}_2)) + i
\theta (U({\bf S}_1)-U({\bf S}_2))} \right]
\end{equation}
The average on the r.h.s. can be readily
performed as the $M$ clauses are drawn independently. The trace on the
two configurations reduces to a sum on the Hamming distance
between them. Evaluation of this sum and the integral over
$\theta$  by the Laplace method yields
\begin{equation}
\underset{\varphi_0}{\mbox{ext}}\left[ f_2(\varphi_0,\alpha) - 2
\alpha x \varphi_0 \right] = S(x,\alpha)
\end{equation}
where $S(x,\alpha)$ is the maximum over $\gamma $ of
\begin{equation}
S(\gamma,x,\alpha) = \ln 2 - \gamma \ln \gamma - (1- \gamma) \ln
(1-\gamma) - \alpha x + \alpha \ln \left[ 1 + p_e(\gamma) (\cosh x -
1) \right]
\end{equation}
and $p_e(\gamma)=(1+(1-2\gamma)^K)/2$ 
is the probability that a randomly  drawn clause
satisfies (or violates) two configurations at
Hamming distance $d=\gamma N$.
We are thus left with the problem of determining $S(x,\alpha)$
and of computing its Legendre transform with respects to $x$ to 
obtain $f_2$. 
As the derivative of
$p_e$ in $\gamma=1/2$ vanishes, this point is always an extremum of
$S$. Two cases must be distinguished,depending on the value of
$\varphi_0$, which fixes $x$\cite{crei2,xorsat2,moore}:
\begin{itemize}
\item If $\gamma=1/2$ is the global maximum of $S$, then
$f_2(\varphi_0,\alpha) = 2 f_1(\varphi_0,\alpha)$, in other words
$[D_U^2] \sim [D_U]^2$. In that case it is possible to compute the
polynomial corrections by expanding around the saddle point, 
\begin{equation}
\frac{[D_U^2]}{[D_U]^2} \sim 1 + \frac{1}{N^{K-2}} \frac{\alpha^2
K!}{2} \left( 1 - 4 \varphi_0 (1- \varphi_0) \right)
\end{equation}
\item If the global maximum of $S$ is not in $\gamma=1/2$, $f_2 >
f_1$, thus $[D_U^2] \gg [D_U]^2$.
\end{itemize}
We have computed numerically for $K=3$ the function
$\varphi_0^{(2)}(\alpha)$ such that for $\varphi_0
>\varphi_0^{(2)}(\alpha)$, the global maxima of $S$ is located in
$\gamma=1/2$. It is a growing function of $\alpha$, vanishing when
$\alpha<0.889$~\cite{crei2}, and growing monotously to $1/2$ when
$\alpha$ diverges. The results are shown in Fig.~\ref{fig_moments}
(all three curves reaches $\varphi _0=1/2$ 
when $\alpha\to\infty$  without crossing each other).
In the course of the algorithm operation, $\varphi _0$ decreases
from its initial value (1/2) down to its plateau value, and remains 
confined to the region in the phase diagram where the second
moment method applies: $\varphi_0 > \varphi_0^{(2)} >
\varphi_0^{(1)}$.  
This proves that, within the Markovian approximation,
the annealed average is correct: as the denominator of the 
transition matrix is peaked
around its mean value, the numerator and denominator can be averaged
separately.
This analysis cannot been done in the case of $K$-SAT, for which the
second moment fails as soon as $\alpha>0$\cite{moore}.

\end{document}